\documentclass{aa}

\usepackage{graphicx}

\usepackage{txfonts}

\usepackage{natbib}
\bibpunct{(}{)}{;}{a}{}{,}

\usepackage{xcolor}
\usepackage{amsmath}
\usepackage{amsfonts}
\usepackage{amssymb}
\usepackage{booktabs}
\usepackage{float}
\usepackage{multirow}
\usepackage[switch]{lineno}

\usepackage[colorlinks=true, linkcolor=blue, citecolor=blue, urlcolor=blue]{hyperref}

\begin{document} 


\title{Investigating the mysterious nature of 1LHAASO J1740+0948u through deep \textit{XMM-Newton} observations}

\author{G. Brunelli\inst{1,2}
      \and      
      G. Ponti\inst{3,4,5}
      \and
      H. Zhang\inst{3}
      \and
      E. de O\~na Wilhelmi\inst{6}
      \and
      V. Sguera\inst{2}
      \and 
      C. Vignali\inst{1,2}
      \and
      R. Zanin\inst{7}
      }

\institute{
    Dipartimento di Fisica e Astronomia (DIFA) ``Augusto Righi”, Università di Bologna, via Gobetti 93/2, I-40129 Bologna, Italy
    \and
        INAF - Osservatorio di Astrofisica e Scienza dello spazio di Bologna, Via Piero Gobetti 93/3, 40129 Bologna, Italy\\
        \email{giulia.brunelli@inaf.it}
    \and
        INAF – Osservatorio Astronomico di Brera, Via E. Bianchi 46, 23807 Merate, Italy       
    \and
        Max-Planck-Institut für extraterrestrische Physik, Gießenbachstraße 1, 85748 Garching, Germany
    \and
        Como Lake Center for Astrophysics (CLAP), DiSAT, Università degli Studi dell’Insubria, via Valleggio 11, 22100 Como, Italy
    \and
        Deutsches Elektronen-Synchrotron DESY, Platanenallee 6, 15738 Zeuthen, Germany
    \and
        Cherenkov Telescope Array Observatory gGmbH, via Gobetti 93, 40129, Bologna, Italy
    }

\date{Received 26/03/2025; accepted 14/10/2025}

\abstract
    {1LHAASO J1740+0948u is a very-high-energy (VHE) source initially reported in the first catalogue by the LHAASO Collaboration, with no previous identifications and no counterpart at other wavelengths. It is detected by the KM2A instrument only, i.e. at energies above 25 TeV, with a $17.1\sigma$ significance, and also above 100 TeV at a $9.4\sigma$ level. It is located ($\sigma_{\rm RA, DEC}\sim0.02^\circ$ at 95\% confidence) at 0.22$^\circ$ from PSR J1740+1000, a faint radio and gamma-ray pulsar placed well above the Galactic plane ($b=20.4^\circ$) that displays a long X-ray tail. Despite the offset, the two sources are likely associated with each other, since no other object has been found nearby at such a high Galactic latitude.}
    {We aim to study the diffuse X-ray emission around PSR J1740+1000 and its tail-like pulsar wind nebula (PWN) with \textit{XMM-Newton} to investigate the origin of the VHE source 1LHAASO J1740+0948u through a multi-wavelength spectral energy distribution (SED) fitting, testing different scenarios.}
    {We analysed $\sim$ 500 ks of \textit{XMM-Newton} observations of PSR J1740+1000. We studied, for the first time, the diffuse emission in two different regions: one centred on the pulsar and the other located inside the 1LHAASO J1740+0948u source region. We also studied the X-ray tail and how its emission evolves as a function of the distance from the pulsar. We then performed a fit of the SED, including the spectrum of 1LHAASO J1740+0948u and the X-ray data obtained from either the analysis of the PWN or the diffuse emission, to understand whether one of the two X-ray sources could be related to the VHE emission and attempt a source classification.}
    {The X-ray analysis of the diffuse emission resulted in upper limits in the range of 0.5--10 keV. The tail-like PWN is best fitted with an absorbed power law with $\Gamma=1.76 \pm 0.06$ in the 0.5--8 keV range, with no significant detection of spectral variations with distance. The SED modelling, assuming the VHE emission to be only due to the X-ray tail, constrains its magnetic field to $B=6.8 \pm 1.9 \ \mu$G, which is in line with previous results. However, we do not find a good fit that could explain both the X-rays of the tail and the LHAASO spectrum with reasonable parameters, hinting that the VHE emission likely comes from an older X-ray-faint electron population. We then performed a SED fitting of the VHE spectrum combined with the upper limits on the diffuse X-ray emission, constraining the magnetic field to be as low as B $\leq1.2\ \mu$G. We suggest that 1LHAASO J1740+0948u could represent either the relic PWN of PSR J1740+1000 or its pulsar halo. Based on our best-fit results, we estimated the energy density and obtained values ranging from 0.03 to 0.67 eV/cm\textsuperscript{3}, depending on the spectral index of the electron distribution. These very low values suggest a halo-like nature for 1LHAASO J1740+0948u, but deeper multi-wavelength observations are required to confirm this hypothesis.}
    {}


\titlerunning{1LHAASO J1740+0948u through XMM-Newton observations}
\authorrunning{G. Brunelli et al.}

\maketitle

\section{Introduction}
Pulsars (PSRs) are one of the most powerful particle accelerators of our Galaxy and give rise to highly energetic phenomena that can be observed up to the very-high-energy (VHE, 50 GeV $\lesssim E \lesssim$ 100 TeV) domain \citep{VERITAS_2011, HESS_2023}. \cite{Goldreich_1969} proved that pulsars are surrounded by the magnetosphere, a region filled with charged particles that can be accelerated with large Lorentz factors through the open magnetic field lines. Later on, \cite{Rees_1974} showed that the accelerated particles escaping the magnetosphere interact with the surrounding medium, producing the pulsar wind nebula (PWN) via synchrotron emission. During their evolution, pulsars, especially older ones, may travel distances greater than the radius of their original PWN, eventually escaping from it. When this occurs, the original PWN begins to evolve independently of the parent pulsar, giving rise to the so-called relic PWN, while the escaped pulsar generates a new PWN \citep{van_der_Swaluw_2004}. In some cases, neutron stars may acquire a kick velocity during the supernova explosion and, when old enough, can also escape the parent supernova remnant (SNR), giving rise to the phenomenon of bow-shock PWNe (BSPWNe). These PWNe are produced by pulsars travelling with a supersonic speed inside the interstellar medium (ISM) after exiting the SNR \citep{Kargaltsev_2017, Olmi_2023}, and they exhibit a variety of different morphologies.

Pulsar wind nebulae are observed from radio to X-rays through their synchrotron emission, and they can generate gamma rays through inverse Compton scattering (ICS) of the ambient photons up to the tera-electronvolt regime and even above 100 TeV, at ultra-high-energies \citep[UHEs;][]{deOnaWilhelmi_2022, Breuhaus_2022}. Relic PWNe are expected to be X-ray faint, because they host a population of cooling old electrons, but still observable at VHEs thanks to ICS emission \citep{deJager_2009}. Many PWNe have been detected at VHEs and UHEs, hence they have been considered one of the possible candidates for the origin of leptonic `PeVatrons', i.e. the Galactic sources responsible for the production of electrons and positrons at energies above 1 PeV. On the other hand, BSPWNe are commonly detected as H$\alpha$ nebulae or in the radio and X-ray bands \citep{Pellizzoni_2004, Kargaltsev_2017}. 

A new class of sources known as pulsar halos (or TeV halos) has been recently established. They were discovered in 2017 \citep{HAWC_2017} as regions of degree-wide tera-electronvolt emission around the Geminga and Monogem PSRs, which could not be attributed to the associated PWNe due to their significantly larger extension of approximately tens of parsecs. They are thought to have originated via ICS between cosmic microwave background (CMB) photons and VHE electrons and positrons accelerated by the central PSR that have escaped into the ISM. Following the discovery, \cite{Giacinti_2020} defined a pulsar halo as the third stage of the evolution of a PSR-PWN system. The first stage is the free expansion of the PWN, lasting until the age of the PSR is$t\lesssim$10 kyr. This is followed by the evolution after the SNR's reverse shock encounters the PWN, lasting up to $t\lesssim$100 kyr. In both stages, the pulsar-originated particles are still confined in the parent SNR, or only a few of them are able to escape, and no pulsar halo can be observed. Only when the PSR becomes middle-aged (i.e. $t\gtrsim100$ kyr), due to its escape from the SNR or to the fading of the SNR itself, are VHE particles free to travel inside the ISM and generate halos. Adopting this definition, halos are identified as systems where the PSR, which generated the relativistic particles, no longer dominates the ISM dynamics or composition anymore \citep{Giacinti_2020}. In particular, the energy density can be used as a potential phenomenological indicator to distinguish between halos and tera-electronvolt PWNe by comparing it to that of the ISM. As a consequence, only the extended emission detected by HAWC around Geminga and Monogem \citep{HAWC_2017} and that detected by LHAASO around PSR J0622+3749 \citep{LHAASO_2021} can be classified as a proper pulsar halo. However, a second definition has been proposed by \cite{Sudoh_2019} and describes a halo as a region hosting relativistic particles where diffusion dominates the particle transport. This definition is independent of the age of the parent PSR and implies that halos could also be found in younger systems. For this reason, many other candidate sources have been proposed in the latest VHE catalogues \citep{Albert_2020, Cao_2024}.

The first catalogue published by the Large High Altitude Air Shower Observatory, LHAASO \citep{Cao_2024}, reported many new unidentified sources detected at both VHEs and UHEs. Of the listed 90 sources, 43 were detected above 4$\sigma$ at energies larger than 100 TeV, 25 were discovered for the first time, and 35 are located in the vicinity of pulsars. 1LHAASO J1740+0948u belongs to all the previous categories. It has been detected only by the KM2A instrument, i.e. above 25 TeV, with a significance of $\sim 12\sigma$ (TS = 156), which decreases to $\sim 6\sigma$ (TS$_{100}$ = 37) above 100 TeV. Recent results by \cite{LHAASO_2025} obtained with additional observation time on the source reported an improved detection at $17.1\sigma$ and $9.4\sigma$ above 25 and 100 TeV, respectively. They derived a 95\% confidence level upper limit on the extension of 0.147$^\circ$ and a log-parabola best-fit spectrum showing a peak at $\sim30$ TeV and extending up to $\sim300$ TeV. The closest pulsar to 1LHAASO J1740+0948u reported in the Australia Telescope National Facility (ATNF) Pulsar Catalogue\footnote{\url{https://www.atnf.csiro.au/research/pulsar/psrcat/}} \citep{Manchester_2005} is PSR J1740+1000, with an angular separation of 0.22$^\circ$. Despite this offset, the absence of other sources near the LHAASO coordinates makes the association with the pulsar likely.

PSR J1740+1000 (J1740 hereafter) is a radio pulsar discovered during a 430 MHz survey of the Arecibo Telescope \citep{McLaughlin_2000} and later detected in X-rays by \textit{Chandra} and \textit{XMM-Newton} \citep{Kargaltsev_2008, Rigoselli_2022}, as well as in the gamma-rays by \textit{Fermi}-LAT \citep{Abdollahi_2022}. It has a period of $P=154$ ms and a period derivative of $\dot{P}=2.1 \cdot 10^{-14}$ s s$^{-1}$, from which the characteristic age ($\tau_C=P/2\dot{P}=114$ kyr) and spin-down power ($\dot{E}=2.3 \cdot 10^{35}$ erg s\textsuperscript{-1}) can be obtained. The distance estimated through the most updated dispersion measure data is $d=1.23$ kpc \citep{Yao_2017}. The source is located significantly above the Galactic plane ($b=20.4^\circ$) and \cite{McLaughlin_2002} suggested a halo origin of the progenitor star, a hypothesis reinforced by the X-ray detection of a long $\sim5-7'$ cometary tail extended in the south-western direction \citep{Kargaltsev_2008}.

Due to its X-ray tail, J1740 can be considered an example of a BSPWN. \cite{Benbow_2021} attempted to detect its tera-electronvolt counterpart, as well as for a few other X-ray BSPWNe, using the Very Energetic Radiation Imaging Telescope Array System (VERITAS) telescopes. In their work, they analysed \textit{XMM-Newton} observations of J1740 in the range 0.3--10 keV and found that the emission is best fitted by a power law with spectral index $\Gamma = 1.75 \pm 0.04$. For the VHE counterpart, instead, only upper limits were obtained. They then developed a model to describe the synchrotron losses of the electron distribution as the particles propagate along the tail, which was assumed to have a cylindrical geometry. They fitted the model to match the X-ray observations and used it to derive the spectral energy distribution (SED) of the corresponding VHE-emitting population. According to their modelling, with improved flux sensitivity of tera-electronvolt instruments, it could become possible to detect a counterpart of the X-ray tail of J1740.

Considering its offset with respect to the pulsar position, \cite{LHAASO_2025} argued that 1LHAASO J1740+0948u may not be the exact counterpart of the X-ray tail, but it could represent the VHE detection of re-accelerated electrons coming from the bow-shock pulsar tail. In particular, they performed a phenomenological model of the synchrotron and ICS losses from a leptonic population and concluded that, for magnetic field values of B = 3, 5 $\mu$G, an X-ray counterpart at the location of 1LHAASO J1740+0948u could be detected with appropriate exposure time. They also explored different scenarios, including a hadronic hypothesis, which was excluded due to the absence of molecular clouds in the surroundings of the pulsar location, and a pulsar halo scenario, which was disfavoured due to the small size and offset of the LHAASO source and J1740. However, they claim that projection effects or very slow diffusion could possibly still be compatible with a pulsar halo origin of the source.

Another study on 1LHAASO J1740+0948u was carried out by \cite{Xie_2025}. They collected multi-wavelength data at the source location, including \textit{Planck}, \textit{Swift}-XRT, \textit{Fermi}-LAT, CfA \textsuperscript{12}CO, and \textit{IceCube} to complement the LHAASO spectrum reported in \cite{Cao_2024}. No counterpart has been found at other wavelengths. They then performed a leptonic SED modelling, assuming the source is powered by a Geminga-like PWN, and derived the main parameters of the relativistic particle population.

Due to their interplay with VHE emission and the better resolution of the instruments when compared to the gamma-ray ground-based facilities, X-rays are an important tool to help us better understand the nature of both PWNe and halos. In particular, since pulsar halos are produced via ICS of the ambient photon fields by the highly relativistic particle wind generated from the pulsar, we would expect to detect synchrotron radiation from the same population of particles and observe it in the X-ray band \citep{Liu_2019, Li_2021}. No counterpart of a pulsar halo in the X-rays has been detected so far, but some attempts have been performed, see for example \cite{Khokhriakova_2024, Manconi_2024, Adams_2025}. In a recent work by \cite{Niu_2025}, the authors claimed the detection using eROSITA of extended X-ray emission around PSR B0656+14, i.e. the Monogem pulsar, but and independent work by \cite{Khokhriakova_2025}, which adopted the most up-to-date eROSITA point spread function (PSF) models, proved that the soft diffuse emission detected by \cite{Niu_2025} can be modelled as the Monogem pulsar emission, suggesting the diffuse emission is due to PSF leakage instead of an additional component.

In this work, we aim to characterise for the first time the diffuse X-ray emission around PSR J1740+1000 using deep \textit{XMM-Newton} observations in order to investigate the presence of a putative counterpart of the VHE source 1LHAASO J1740+0948u. We also revisit the analysis of the tail-like PWN, performing the first study of the spectral index as a function of the distance from the pulsar. We then perform a multi-wavelength SED fitting, using both the X-ray fluxes derived from the spectral analysis of the tail and the upper limits obtained from the diffuse emission. We also used the newly available tera-electronvolt spectrum from \cite{LHAASO_2025} to test different possible acceleration scenarios that may explain the nature of the dark VHE source. The paper is divided as follows: in Sect. \ref{sec:xr_analysis} we present the methodology followed for the X-ray analysis of both the diffuse emission and the tail PWN, in Sect. \ref{sec:results} we show the derived results and the SED fitting, and in Sect. \ref{sec:discussion} we discuss the results interpretation.

\section{X-ray observations and data analysis} \label{sec:xr_analysis}
The European Photon Imaging Cameras (EPIC) on board \textit{XMM-Newton} observed PSR J1740 between 2017 and 2018 for a total of five observations (ObsIDs: 0803080101 - 0803080501) and twice in 2006 (ObsIDs: 0403570101, 0403570201). We focused on the 2018 sample due to the large total exposure time compared to the 2006 sample. Out of the five observations, the first one (ObsID: 0803080101) was not considered due to both EPIC-MOS1/2 and EPIC-pn instruments being affected by high background from solar flares. The exposure time for the four observations together reaches a total of 532.5 ks. The EPIC-pn instrument was operated in `Small Window' mode, i.e. with only a part of CCD4 used to collect data, while both EPIC-MOS1 and MOS2 were operated in `Full Frame' mode, i.e. all CCDs in read-out mode \citep{xmm_handbook}. Moreover, CCD3 and CCD6 of the MOS1 have not been operational since 2005 and 2012, respectively, due to impact with micrometeorites \citep{xmm_handbook}. For these reasons, in the analysis of the diffuse emission around the pulsar, we used only data collected by the MOS2 camera that recorded the full field of view (FoV) available to the camera. The same holds for the analysis of the tail, which, for some observations, was not fully covered by the available MOS1 chips. We performed the data reduction using version \texttt{v21} of the \texttt{Science Analysis System} (\texttt{SAS})\footnote{\url{https://www.cosmos.esa.int/web/xmm-newton/sas}}, but used two different approaches for the analysis of the diffuse emission and the PWN.

\subsection{Analysis of the diffuse emission} \label{sec:diffuse_analysis}
To study the diffuse emission, we employed the \texttt{XMM-Newton Extended Source Analysis Software} (\texttt{XMM-ESAS}) package\footnote{\url{https://www.cosmos.esa.int/web/xmm-newton/xmm-esas}} available in \texttt{SAS}. The pipeline we followed for the analysis started with the raw data reduction with \texttt{emchain}. Then, we filtered for the soft proton flares using the \texttt{histogram} method for the task \texttt{espfilt}, setting a flare-free region at 3$\sigma$ around the peak of the Gaussian distribution of the count rates. The next step was to run the task \texttt{cheese} for the source detection in the entire FoV of MOS2, for which we used a minimum maximum likelihood detection value of \texttt{mlmin=5} (corresponding to a significance of $\sim4\sigma$). This task produced a mask for the selected background sources that was later used in the spectral extraction phase.

\begin{figure*}[ht]
    \centering
    \includegraphics[width=\linewidth]{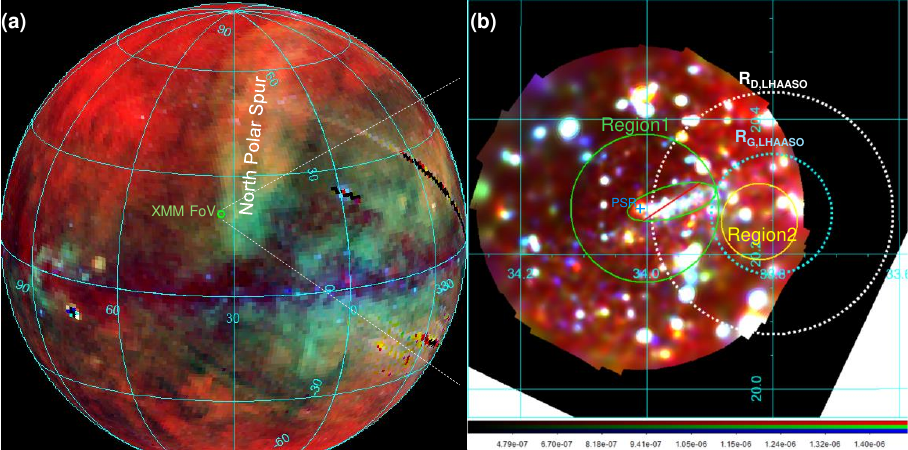}
    \caption{\textit{(a)} ROSAT all-sky map showing the large-scale diffuse emission of the North Polar Spur and the location of the \textit{XMM-Newton} pointings, i.e. the small green circle. The red, green and blue colours represent the 0.11--0.28 keV, 0.47--1.21 keV and 0.76--2.04 keV energy bands, respectively. \textit{(b)} Image of the \textit{XMM-Newton} FoV, showing both Region 1 and Region 2 (solid circles), and 1LHAASO J1740+0948u (dashed circles). The plotted radii of the source are taken from the 1$\sigma$ extension of the Gaussian ($R_{G, LHAASO}$) and the diffuse ($R_{D, LHAASO}$) models for photon energy above 25 TeV, shown in \cite{LHAASO_2025}. The 0.5--1 keV band is shown in red, the green emission corresponds to the range 1--2 keV, and the range 2--4.5 keV is represented in blue. Note that the observed point sources were later removed during the diffuse emission analysis.}
    \label{fig:fov_xmm}
\end{figure*}

The goal of this part of the analysis was to detect diffuse non-thermal X-ray emission potentially associated with 1LHAASO J1740+0948u. The main challenge we encountered when selecting the source region was that the full extension of 1LHAASO J1740+0948u is not covered by the \textit{XMM-Newton} FoV. For this reason, we investigated two regions. The first, dubbed `Region 1' hereafter, was selected as a 0.11$^\circ$ circle centred on the pulsar and excluding an ellipse comprising both the pulsar and the PWN. This region covers part of the extension of the LHAASO source and approximately the entire area of the central CCD, where the PSF of \textit{XMM-Newton} has minimal distortions when compared to the on-axis PSF. The second, `Region 2' from now on in the paper, was selected as a circular $r=0.06^\circ$ region in the outer part of the XMM FoV, corresponding to the majority of the 68\% source extension reported by \cite{LHAASO_2025}. Both regions are shown in panel (b) of Fig. \ref{fig:fov_xmm}, along with the extension of 1LHAASO J1740+0948u.

Finally, we performed the spectral extraction using \texttt{mosspectra} in the energy range 0.5 - 10 keV. This task also generated the corresponding response matrix file (RMF) and ancillary response file (ARF). Note that we did not run the task \texttt{mosback} because we wanted to fit the background, both instrumental and physical components, with a custom-made model. We adopted this approach as our goal was to investigate the presence of an extended component that may extend beyond the FoV of the instrument.

The whole procedure was repeated for each observation individually. Once the final spectrum with its RMF and ARF were obtained for all four observations, we stacked them to obtain the total spectrum. To do so, we used some of the tasks of the \texttt{ftools} package: \texttt{mathpha}, to stack the source spectra, and \texttt{addrmf} and \texttt{addarf} to produce an average RMF and ARF. Computing an average response is necessary to take into account the differences between the individual pointings. The final source spectrum was grouped using \texttt{grppha} with a minimum of 200 counts per bin. The spectral modelling and fitting of both the source and background were performed with version \texttt{v12.13.1} of the \texttt{Xspec} package \citep{Arnaud_1996}.

\subsubsection{Background modelling} \label{sec:back_model}
The instrumental background of \textit{XMM-Newton} consists of several components: the quiescent particle background (QPB), the contamination by soft proton flares (SPF) and the contamination by solar wind charge exchange (SWCX). The first was already treated during the data reduction, so we focused on the QPB component, that can be described as a combination of a flat continuum emission and instrumental lines. We followed the prescriptions of \cite{Leccardi_Molendi_2008} and modelled the continuum as a broken power law (\texttt{bknpower}) with $\Gamma_1=0.32$, $\Gamma_2=0.22$ and $E_b=3$ keV. Being this component intrinsic to the instrument, we did not convolve it with the ARF. We performed the fit by leaving all the parameters of the broken power law, except the normalisation, free. We then added the instrumental lines as Gaussian components (\texttt{gauss}) one by one to verify whether the new component was significant. Out of all the lines listed in \cite{Leccardi_Molendi_2008} we included: Al K$\alpha$, Si K$\alpha$, Cr K$\alpha$, Mn K$\alpha$, Cr K$\beta$, Fe K$\alpha$, Ni K$\alpha$ and Au L$\alpha$. We left the energy and normalisation of all lines free to vary and fixed the width to zero for all except the Al K$\alpha$ and Si K$\alpha$. We also verified the presence of SWCX contamination in the dataset. We found no significant flux enhancement in the soft band among the four observations, and the fluxes of all the background components are compatible, considering the statistical uncertainties.

The physical background, on the other hand, is not intrinsic to the instrument and can change depending on the observation and the pointing. The model we used comprised several components:
\begin{enumerate}
    \item Unabsorbed thermal component representing the local hot bubble (LHB) with fixed kT = 0.097 keV and Z = $Z_{\odot}$ \citep{Ponti_2023} and free normalisation.
    \item Absorbed thermal component representing the Galactic halo emission, with fixed Z = $0.1Z_{\odot}$ and free temperature and normalisation.
    \item Absorbed thermal component representing the North Polar Spur (NPS), a region of enhanced soft X-ray emission that is part of the radio Loop I \citep{Kataoka_2018}. The projected position of J1740 is on the edge of the NPS, as can be seen from Fig. \ref{fig:fov_xmm}, so its emission cannot be neglected for the background modelling. We fitted it with all the parameters free to vary because fixing a temperature and an abundance would be too restrictive, since the NPS is probably a multi-component plasma.
    \item An absorbed double broken power law representing the cosmic X-ray background (CXB), following the method of \cite{Ponti_2023}, in turn based on the model of \cite{Gilli_2007}. We let the normalisation free to vary, but it was later fixed to avoid a degeneracy with the power law component later added to model the source (see Sec. \ref{sec:halo_model}).
\end{enumerate}
For the absorption, we first employed the \texttt{tbabs} model described in \cite{Wilms_2000} and fixed column density $n_X$ to the value at the coordinates of the pulsar, $n_X=8 \cdot 10^{20}$ cm$^{-2}$, derived with the \texttt{Xspec} task \texttt{nh}. Then, to properly take into account the possible fluctuations of the column density inside the studied regions, we substituted \texttt{tbabs} with the \texttt{disnht} model developed by \cite{Locatelli_2022}. This model assumes the column density to be distributed as a log-normal function and can be considered as a generalisation of \texttt{tbabs}, better suited for extended and faint sources. In the end, the expression for the model used in the fit of the physical background is \texttt{apec$_\texttt{LHB}$ + disnht*(apec$_{\texttt{Halo}}$ + apec$_\texttt{NPS}$ + bkn2pow$_\texttt{CXB}$)}.

\subsubsection{Source modelling} \label{sec:halo_model}
To search for diffuse non-thermal X-ray emission in the two regions, we added an absorbed power law (\texttt{disnht*powerlaw}) component to the model, leaving both normalisation and index free to vary. The parameters of \texttt{disnht} were set to be equal to those of the physical background model. To avoid a degeneracy with the CXB component, we fixed its normalisation to the expected value for the area considered in the source region. To do so, we used the best-fit value of the model by \cite{Gilli_2007} and derived the normalisation for the areas of both Region 1 and Region 2. The results are $n_{CXB}=8.28 \cdot 10^{-5}$ ph $\cdot$ s $\cdot$ cm\textsuperscript{2} and $n_{CXB}=2.57 \cdot 10^{-5}$ ph $\cdot$ s $\cdot$ cm\textsuperscript{2} for Region 1 and 2, respectively. We then performed the fit in the range 0.5 keV--10 keV. 

\subsection{Analysis of the tail-like PWN}
\begin{figure}[ht]
    \centering
    \includegraphics[width=\linewidth]{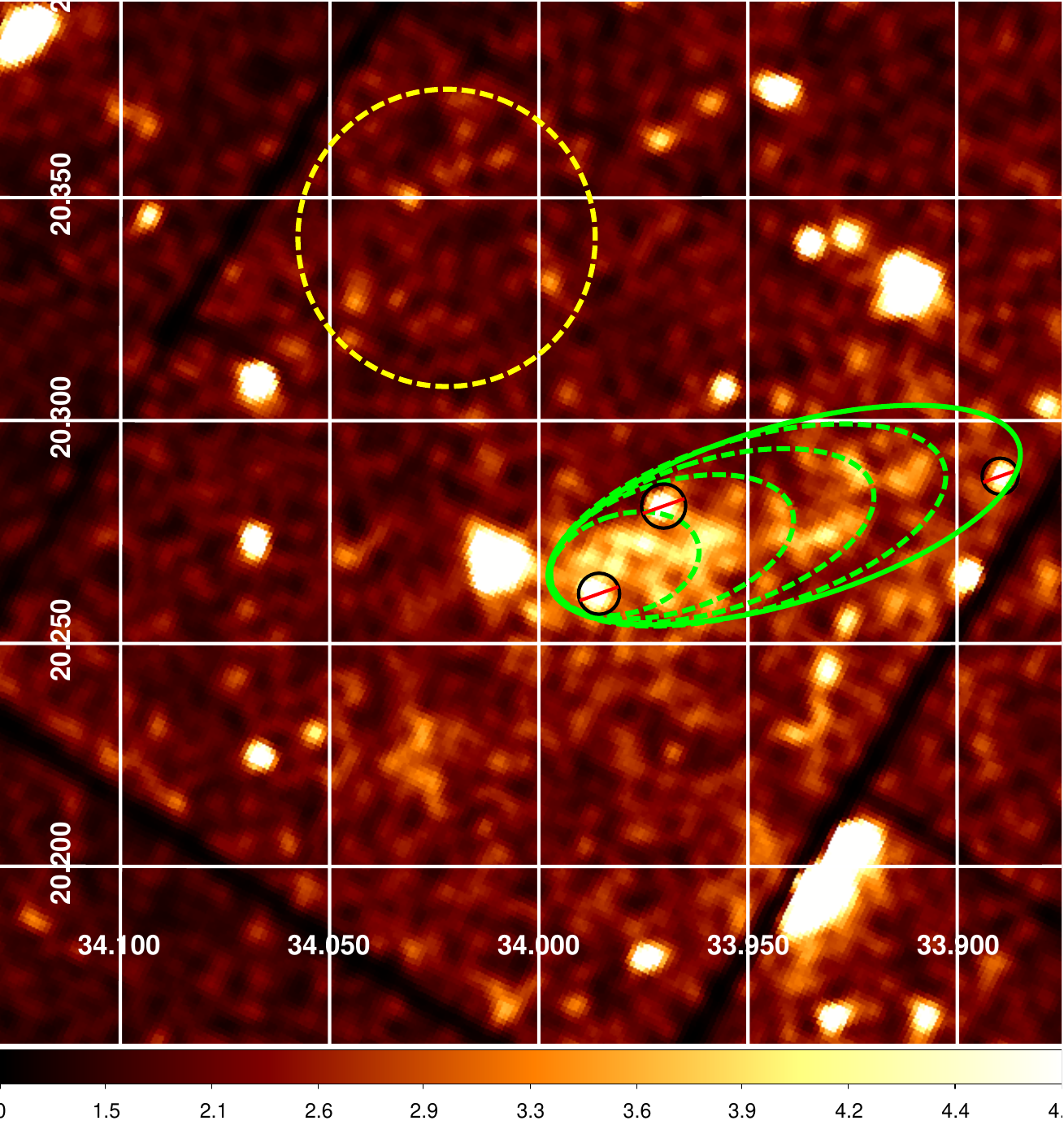}
    \caption{\textit{XMM-Newton} image of the tail region in the energy range 0.5--8 keV. The solid ellipse represents the region where the overall tail emission has been extracted, while the dashed ellipses mark the borders of the sub-tail regions. The dashed circle is the background extraction region. We also display the Galactic coordinates grid.}
    \label{fig:tails_and_back}
\end{figure}

As for the diffuse emission analysis, we employed MOS2 data only and reduced them using the task \texttt{emproc} to obtain the event files. We lowered the contribution of the SPF by selecting the good time intervals after applying a maximum limit on the count rates to the lightcurve above 10 keV. For the source spectral extraction, we selected the elliptical region shown in Fig. \ref{fig:tails_and_back} (solid ellipse), excluding the three point sources (small black circles) displayed in the same figure. The background was chosen as a circular region located in the same CCD but sufficiently far from the pulsar and its PWN to exclude possible contamination from both, and did not include any significant point-like source present in the FoV (see dashed circle in Fig. \ref{fig:tails_and_back}). Even though the background region is included in the extension of Region 1, the contamination by X-ray emission possibly associated with the LHAASO source can be considered negligible, since no excess of non-thermal X-rays was found by performing the analysis of the diffuse emission (see Sect. \ref{subsec:results_diff}). The task \texttt{backscale} was used on both the source and background spectra to properly take into account the different areas of the two regions and the possible chip gaps or bad pixels. The RMF and ARF were produced using the \texttt{rmfgen} and \texttt{arfgen} tasks, respectively. Also in this case, we stacked the four observations by using \texttt{mathpha} on the spectra of both the source and background, and we produced an average RMF and ARF with \texttt{addrmf} and \texttt{addarf}. The final source spectrum was grouped using \texttt{grppha} with a minimum of 175 counts per bin to ensure at least 25 counts per bin in the background-subtracted spectrum.

Finally, we used \texttt{Xspec} to model the tail emission as an absorbed power law \texttt{tbabs*powerlaw}, where we fixed the column density to the value at the pulsar's coordinates $n_X=8 \cdot 10^{20}$ cm$^{-2}$ (see Sect. \ref{sec:back_model}). In this case, we employed the simpler \texttt{tbabs} model, since the source is not as extended as the diffuse emission. The fit was performed between 0.5 keV and 8 keV, after which the instrumental background of the CCDs becomes dominant over the physical signal, ignoring the 1.3--1.8 keV band. Due to the choice of the background region, located on one of the edges of the central CCD, some residuals around the Al K$\alpha$ (1.49 keV) - Si K$\alpha$ (1.74 keV) complex could still be observed after the background subtraction. This occurs because of the peculiar distribution of the lines' strength in the detector plane, as shown in \cite{Kuntz_Snowden_2008}. We produced the same plots with the most up-to-date merged filter wheel closed (FWC)\footnote{\url{https://www.cosmos.esa.int/web/xmm-newton/filter-closed}} data. The same line strength distribution of both the Al K$\alpha$ and Si K$\alpha$ reported in \cite{Kuntz_Snowden_2008} is observed.

The tail region previously defined was subsequently divided into five smaller and concentric elliptical regions, as shown in Fig. \ref{fig:tails_and_back} (dashed ellipses). The procedure described above was repeated on each one of them to investigate the evolution of the spectral model as a function of the distance from the pulsar. We fitted the smaller tails in the range 0.5 keV--6 keV, again excluding the 1.3--1.8 keV band, to remove the last energy bins where the minimum of 25 counts in the background-subtracted spectrum was not satisfied. The restriction does not affect the final result, since most of the tail emission is mostly concentrated where the instrumental background is not dominating.

\section{Results of the X-ray analysis} \label{sec:results}
\subsection{Diffuse emission} \label{subsec:results_diff}
When producing the model including only the background components, we first compared the best-fit results of the CXB normalisation to that expected from \cite{Gilli_2007}, $n_{CXB}=8.28 \cdot 10^{-5}$ ph $\cdot$ s $\cdot$ cm\textsuperscript{2} and $n_{CXB}=2.57 \cdot 10^{-5}$ ph $\cdot$ s $\cdot$ cm\textsuperscript{2} for Region 1 and 2, respectively (see Sect. \ref{sec:halo_model}). In the case of Region 1, the best-fit result is $n_{CXB}=(8.2 \pm 0.4) \cdot 10^{-5}$ ph $\cdot$ s $\cdot$ cm\textsuperscript{2}, with errors at 90\% confidence level. The result is fully compatible with the theoretical prediction, suggesting that no additional components are probably needed in the fit. Instead, for Region 2, the best-fit normalisation is $n_{CXB}=(3.9 \pm 0.4)\cdot 10^{-5}$ ph $\cdot$ s $\cdot$ cm\textsuperscript{2}, with 90\% confidence level uncertainties. The value is higher than the expectations, although we note that it is reported without an associated error. This hints that some residual non-thermal emission could likely be present in the region.

As explained in Sect. \ref{sec:halo_model}, we subsequently fitted the diffuse emission with an absorbed power law with free index and normalisation. In the case of Region 1, the non-thermal component was not detected, and its addition did not improve the $\chi^2$ value. This is compatible with what we observed when studying the normalisation of the CXB. In the case of Region 2, instead, the best-fit parameters are $\Gamma=2.1^{+1.9}_{-1.1}$ and $N_0=2.8^{+3.1}_{-1.9} \cdot 10^{-5}$ ph $\cdot$ keV$^{-1} \cdot$ s$^{-1} \cdot$ cm$^{-2}$, with statistical uncertainties at 3$\sigma$ level. The improvement in the $\chi^2$ after adding the absorbed power law is from $\chi^2=240.6$ (178 d.o.f.) to $\chi^2=235.8$ (176 d.o.f.). Even though it could be considered a detection at 3$\sigma$, we treat the result as an upper limit for several reasons. First, we see that the improvement in the $\chi^2$ after adding the source component is not very significant, hinting at a weak source detection. More importantly, some assumptions were made when building the background model, and fluctuations in the choice of the parameters for the adopted functions could easily lead to artefacts that can be observed as true detections. As shown in Fig. \ref{fig:fov_xmm}, there does not seem to be any strong flux in Region 1, while the emission from Region 2 might be affected by the presence of the NPS or fluctuations in the background.

Since we do not have a detection of the diffuse X-ray emission, the spectral index is not determined, so we derived the total upper limit on the X-ray emission by considering different values of the photon index in the spectrum. We obtained the 3$\sigma$ upper limit on the normalisation at 1 keV for values of the photon index between $\Gamma=3.0$ and $\Gamma=0.5$. We represented all the derived power laws in the same plot, and we computed an overall upper limit in 0.5--10 keV. We computed it for both Region 1 and 2, and we subsequently used them together with the spectrum reported in the LHAASO catalogue to perform the SED fitting, see Sect. \ref{sec:sed}. For the SED, we assumed that the emission is uniform as a function of radius. Both upper limits for Region 1 (radius $r_1=0.11^\circ$) and Region 2 (radius $r_2=0.06^\circ$) were rescaled to the area of the LHAASO region, assuming $r_{\rm LHAASO}=0.147^\circ$, as provided by \cite{LHAASO_2025}.

\subsection{Tail-like PWN analysis}
We obtained a good spectral fit ($\chi^2=70.57$ for 75 d.o.f.) of the tail emission for a spectral index of $\Gamma=1.76 \pm 0.06$ and normalisation at 1 keV of $N_0=(3.49 \pm 0.14) \cdot 10^{-5}$ ph $\cdot$ keV$^{-1} \cdot$ s$^{-1} \cdot$ cm$^{-2}$ or $N_0=(33.1 \pm 1.3)$ ph $\cdot$ keV$^{-1} \cdot$ s$^{-1} \cdot$ cm$^{-2} \cdot$ sr$^{-1}$ when expressing it in surface brightness after setting the proper value of the \texttt{AREASCAL} keyword. The 0.5--8 keV unabsorbed flux is $F_x=(1.86 \pm 0.12) \cdot 10^{-13}$ erg $\cdot$ s$^{-1} \ \cdot$ cm$^{-2}$, corresponding to an unabsorbed luminosity of $L_x=(3.4 \pm 0.2) \cdot 10^{31}$ erg $\cdot$ s$^{-1}$ assuming a distance of $d=1.23$ kpc \citep{Yao_2017}. From the luminosity, we derived the X-ray efficiency as $\eta=L_x/\dot{E}=1.5 \cdot 10^{-4}$. The result for the flux is compatible within the statistical uncertainties with that of \cite{Benbow_2021}, $F_X=(1.93 \pm 0.06)\cdot 10^{-13}$ erg $\cdot$ s$^{-1} \ \cdot$ cm$^{-2}$, even though they reported the absorbed flux of the PWN in the energy range 0.3--10 keV. On the other hand, there is a discrepancy between our result for the luminosity and theirs, $L_X=(5.20 \pm 0.16) \cdot 10^{31}$ erg $\cdot$ s$^{-1}$, most likely due to different source and background extraction regions and energy bands employed in the two analyses and a different assumed value for the pulsar distance. Finally, we extracted the SED flux points for this dataset directly from \texttt{Xspec} and used them together with the LHAASO power law spectrum in the SED fitting, also comparing them to the upper limit by \cite{Benbow_2021} (see Sect. \ref{sec:sed}).

\begin{figure}[ht]
    \centering
    \includegraphics[width=\linewidth]{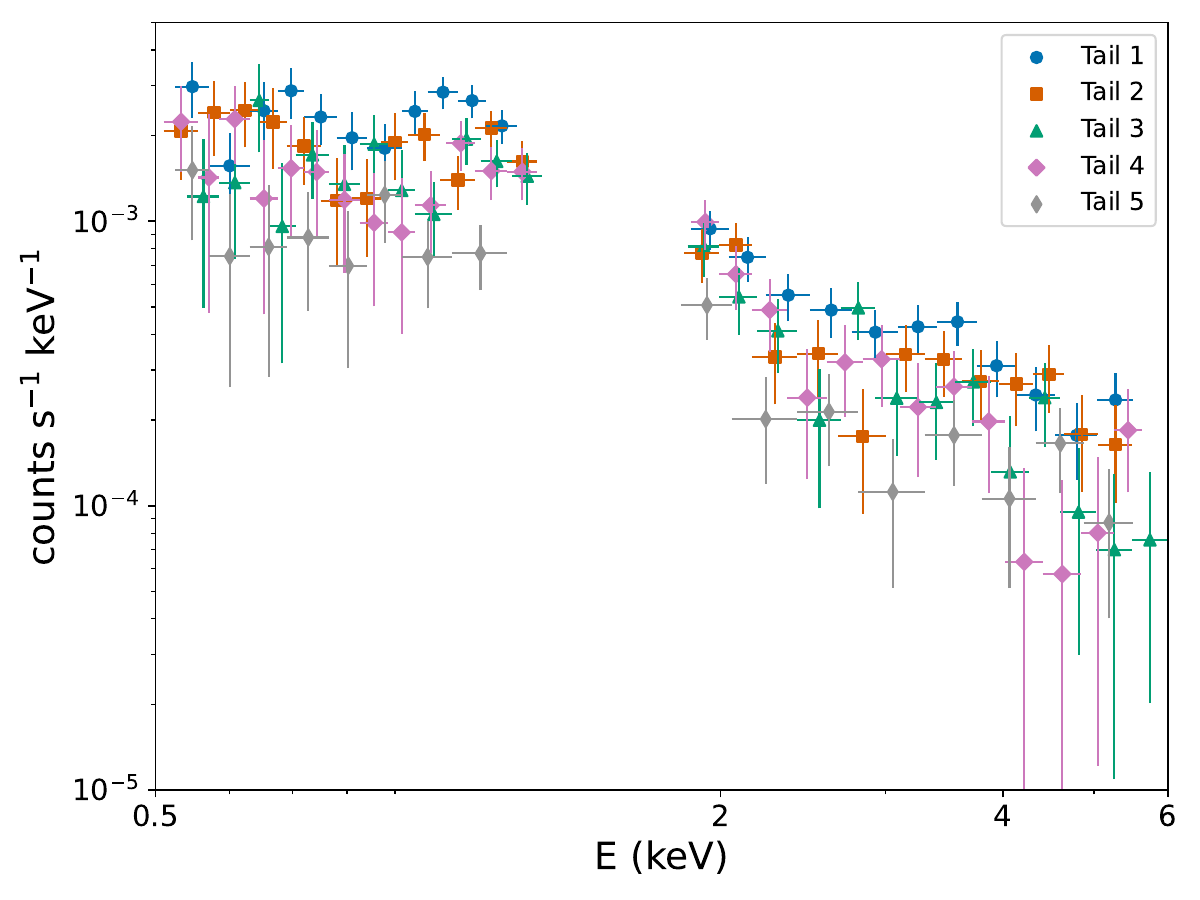}
    \caption{Spectra of the single tails in the energy range 0.5--6 keV, excluding the 1.3--1.8 keV band due to the contamination of instrumental lines. Tail 1 is the innermost with respect to the pulsar, Tail 5 is the outermost.}
    \label{fig:tails_spectra}
\end{figure}

\begin{figure*}[ht]
    \centering
    \sidecaption
    \includegraphics[width=0.6\linewidth]{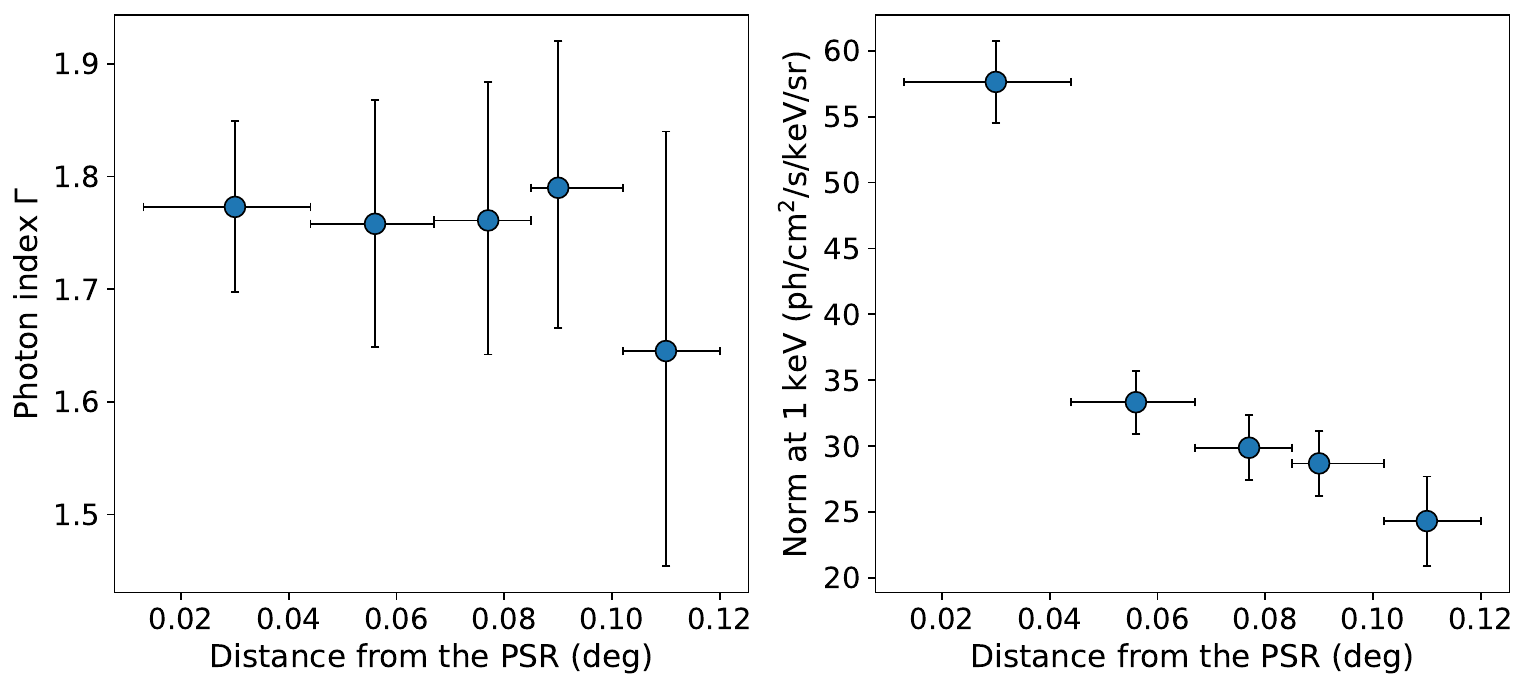}
    \caption{Evolution of the spectral parameters as a function of the angular distance from the pulsar position. The uncertainties on the index and surface brightness values are shown at 1$\sigma$ level. \textit{Left panel}: the trend of the spectral index. \textit{Right panel}: the trend of the normalisation of the power law at 1 keV considering the \texttt{AREASCAL} parameter.}
    \label{fig:tails_with_dist}
\end{figure*}

The same fit procedure has been applied to the single sub-tails, and the results are summarised in Tab. \ref{tab:tails_results}, along with the best-fit values of the model for the overall tail emission. The reported normalisation values are computed by including the region area in the \texttt{AREASCAL} parameter to obtain a surface brightness value at 1 keV. The spectra of the sub-tails are also depicted in Fig. \ref{fig:tails_spectra}. No evidence for an evolution of the spectral index is found, i.e. no synchrotron cooling is detected across the length of the tail, as it can also be observed in Fig. \ref{fig:tails_with_dist} (left). The trend of the surface brightness at 1 keV, illustrated on the right plot of Fig. \ref{fig:tails_with_dist}, shows that the innermost part of the tail, closer to the pulsar, is brighter than the outer part. The horizontal bars associated with the distances correspond to the inner and outer boundaries of the ellipses.

\begin{table*}[ht]
    \centering
    \caption{Best-fit results for the analysis of the overall tail and the five sub-tails shown in Fig. \ref{fig:tails_and_back}.}
    \label{tab:tails_results}
        \begin{tabular}{ccccc}
            \toprule
            \toprule
             & Index & \begin{tabular}[c]{@{}c@{}}Normalisation\\ ($10^{-6}$ ph $\cdot$ keV$^{-1}$ $\cdot$ s$^{-1}$ $\cdot$ cm$^{-2})$\end{tabular} & \begin{tabular}[c]{@{}c@{}}Area normalisation\\ (ph $\cdot$ keV$^{-1}$ $\cdot$ s$^{-1}$ $\cdot$ cm$^{-2}$ $\cdot$ sr$^{-1}$)\end{tabular} & $\chi^2$ (d.o.f.) \\
            \midrule
            Total & 1.76 $\pm$ 0.06 & 3.49 $\pm$ 0.14 & 33.1 $\pm$ 1.3 & 70.57 (75) \\
            \midrule
            Tail 1 & 1.77 $\pm$ 0.08 & 9.5 $\pm$ 0.5 & 58 $\pm$ 3 & 20.78 (20)  \\
            Tail 2 & 1.76 $\pm$ 0.11 & 7.2 $\pm$ 0.5 & 33 $\pm$ 2 & 29.92 (22)  \\
            Tail 3 & 1.76 $\pm$ 0.12 & 6.5 $\pm$ 0.5 & 30 $\pm$ 2 & 17.32 (23)  \\
            Tail 4 & 1.79 $\pm$ 0.13 & 6.8 $\pm$ 0.6 & 29 $\pm$ 2 & 20.37 (24)  \\
            Tail 5 & 1.6 $\pm$ 0.2   & 4.3 $\pm$ 0.6 & 24 $\pm$ 3 & 10.07 (14)  \\
            \bottomrule
        \end{tabular}
        \tablefoot{Tail 1 represents the innermost region and Tail 5 the outermost. We report the spectral index, along with both the flux normalisation at 1 keV and the normalisation obtained by dividing the spectrum by the \texttt{AREASCAL} parameter. We also show the value of the $\chi^2$ of the fit and the number of degrees of freedom. The uncertainties are computed at 1$\sigma$ level.}
\end{table*}

\subsection{SED modelling} \label{sec:sed}
We performed an SED modelling over the X-ray and tera-electronvolt range using the package \texttt{naima} \citep{naima}, which allows the user to study the radiative losses of a relativistic population of particles, considering a purely leptonic scenario. We did not treat a hadronic scenario because it is beyond the aim of this paper. Moreover, \cite{LHAASO_2025} already discussed the low likelihood of this kind of model for 1LHAASO J1740+0948u due to the absence, in the vicinity of the source, of molecular clouds that could give rise to hadronic emission. The goal of this SED modelling was to try to put constraints on the origin of the emission observed by LHAASO by exploring different scenarios.

\subsubsection{Scenario 1: potential VHE detection of the X-ray tail}
We first investigated the hypothesis that 1LHAASO J1740+0948u is the counterpart of the X-ray tail. To this aim, we performed the SED modelling using the previously produced flux points in the range 0.5--8 keV and the KM2A spectrum of \cite{LHAASO_2025} as well as their extrapolation for the WCDA instrument. We assumed the electron distribution at injection to be an exponential cut-off power law (ECPL)
\begin{equation} \label{eq:ecpl}
    f(E)=A \bigg( \frac{E}{E_0} \bigg)^{-p} \exp \bigg[ - \bigg( \frac{E}{E_{cut}} \bigg)^\beta \bigg] ,
\end{equation}
where $E_0$ is the reference energy, fixed to $E_0=1$ TeV, $p$ is the spectral index, $E_{cut}$ is the cut-off energy, $\beta$ the cut-off parameter, fixed to $\beta=2$ as in \cite{Zirakashvili_2007}, and $N_0$ is the normalisation, which is related to the total energy content $W_e$. The minimum and maximum energy of the particle distribution were set to $E_{min}=1$ GeV and $E_{max}=960$ TeV, respectively. The maximum energy was obtained from Eq. (3) of \cite{deOnaWilhelmi_2022}, setting $\eta_e=\eta_B=1$, where $\eta_e$ represents the ratio between the electric field strength and the magnetic one, and $\eta_B$ represents the fraction of pulsar wind energy flux that goes into magnetic energy density \citep{deOnaWilhelmi_2022}. The minimum, instead, was chosen taking into account that no points below the X-ray band are included in the fit. We considered synchrotron and ICS as the emission mechanisms for the particle population. For the latter we included as photon field seeds the CMB, far-infrared (FIR) radiation with $T_{FIR}=33$ K and $u_{FIR}=0.2$ eV/cm$^3$, near-infrared (NIR) radiation with $T_{NIR}=350$ K and $u_{NIR}=0.02$ eV/cm$^3$, and the starlight contribution with $T_{star}=4800$ K and $u_{star}=0.3$ eV/cm$^3$, where all the values were obtained from \cite{Popescu_2017} at the pulsar location.

\begin{figure}[ht]
    \centering
    \includegraphics[width=\linewidth]{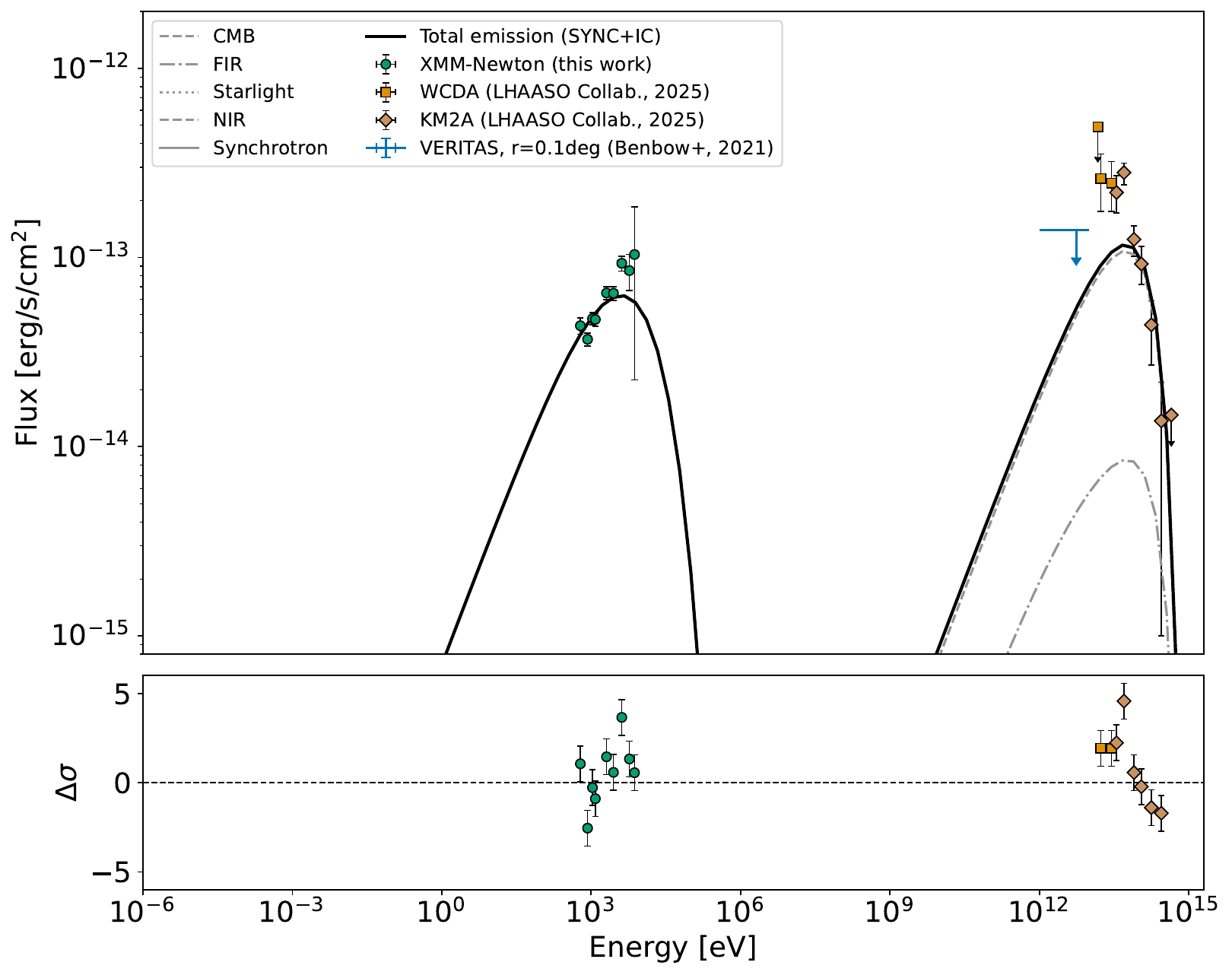}
    \caption{SED of the tail-like PWN of J1740, including the X-ray points derived in this work and the LHAASO-KM2A spectrum along with the LHAASO-WCDA extrapolation of \cite{LHAASO_2025}. The electron distribution is modelled as an ECPL with fixed $p=1.6$, $\beta=2$, $E_0=1$ TeV and free $E_{cut}$ and normalisation. We included ICS, considering all the photon fields obtained from \cite{Popescu_2017}, and Synchrotron, where the magnetic field value was left free to vary. We also report the VERITAS upper limit of \cite{Benbow_2021} obtained for an extraction region of 0.1$^\circ$. The model residuals, obtained as (data - model)/error, are shown in the bottom panel.}
    \label{fig:pwn_SED}
\end{figure}

We first performed the fit, fixing the spectral index to the expectation value from the X-ray spectrum $p=2\Gamma_x -1$ \citep{Kargaltsev_Pavlov_2010} that corresponds to $p=2.52$ for $\Gamma_x=1.76$. The derived best fit model, with parameters $E_{cut}=329^{+15}_{-27}$ TeV, $B=6.8\pm1.9 \ \mu$G and $W_e=1.3\cdot 10^{45}$ erg, only matched the X-ray points and was not able to reproduce the ICS component observed by LHAASO. The magnetic field obtained through this fit, however, is compatible with previous literature results \citep{Kargaltsev_2008, Benbow_2021}. To better fit both X-ray and gamma-ray data, we performed several tests changing the spectral index of the electron distribution. We compared the results of each test by using the Bayesian information criterion, BIC \citep{Schwarz_1978}, defined as $BIC=k\ln(n)-2\ln(\mathcal{L})$, where $k$ is the number of free parameters, $n$ is the number of points in the sample and $\mathcal{L}$ represent the likelihood of the model. The BIC is computed by \texttt{naima} during the fitting procedure, along with the maximum logarithmic likelihood value. Lower values of the BIC usually indicate a statistical preference for the associated model with respect to the other. The model that returned the lowest value of BIC and also had a match with the LHAASO spectrum was the one with $p=1.6$, with best-fit results $E_{cut}=304_{-21}^{+32}$ TeV, $B=1.31^{+0.10}_{-0.07} \ \mu$G and $W_e=1.34 \cdot 10^{44}$ erg. The best-fit SED and its residuals are depicted in Fig. \ref{fig:pwn_SED}. The electron energy distribution from the best-fit model shows a harder spectrum ($p=1.6$) compared to the expected value $p=2\Gamma_x-1=2.52$, and the expected flux at tens of tera-electronvolts is below the observed spectrum by a factor of two or three. Moreover, the best-fit magnetic field value is significantly smaller than that obtained for $p=2.52$ and previous literature results. Even though with \texttt{naima} we are estimating only the perpendicular component of the magnetic field, it has been proven that the total magnetic field is dominated by that component \citep{Liu_Yan_2019}, so the real value of the magnetic field will still be lower than the expected $\sim6 \ \mu$G. This hints that the ICS component generated by the electrons responsible for the X-ray emission is insufficient to explain the tera-electronvolt flux detected by LHAASO. Therefore, it suggests that LHAASO probably did not detect the counterpart of the X-ray tail and that the two sources are related to different populations of electrons, most likely a young electron population generating the X-ray emission and an older population emitting at tera-electronvolt energies.

\subsubsection{Scenario 2: diffuse X-ray emission from the LHAASO source}
\begin{figure*}[ht]
    \centering
    \includegraphics[width=0.49\linewidth]{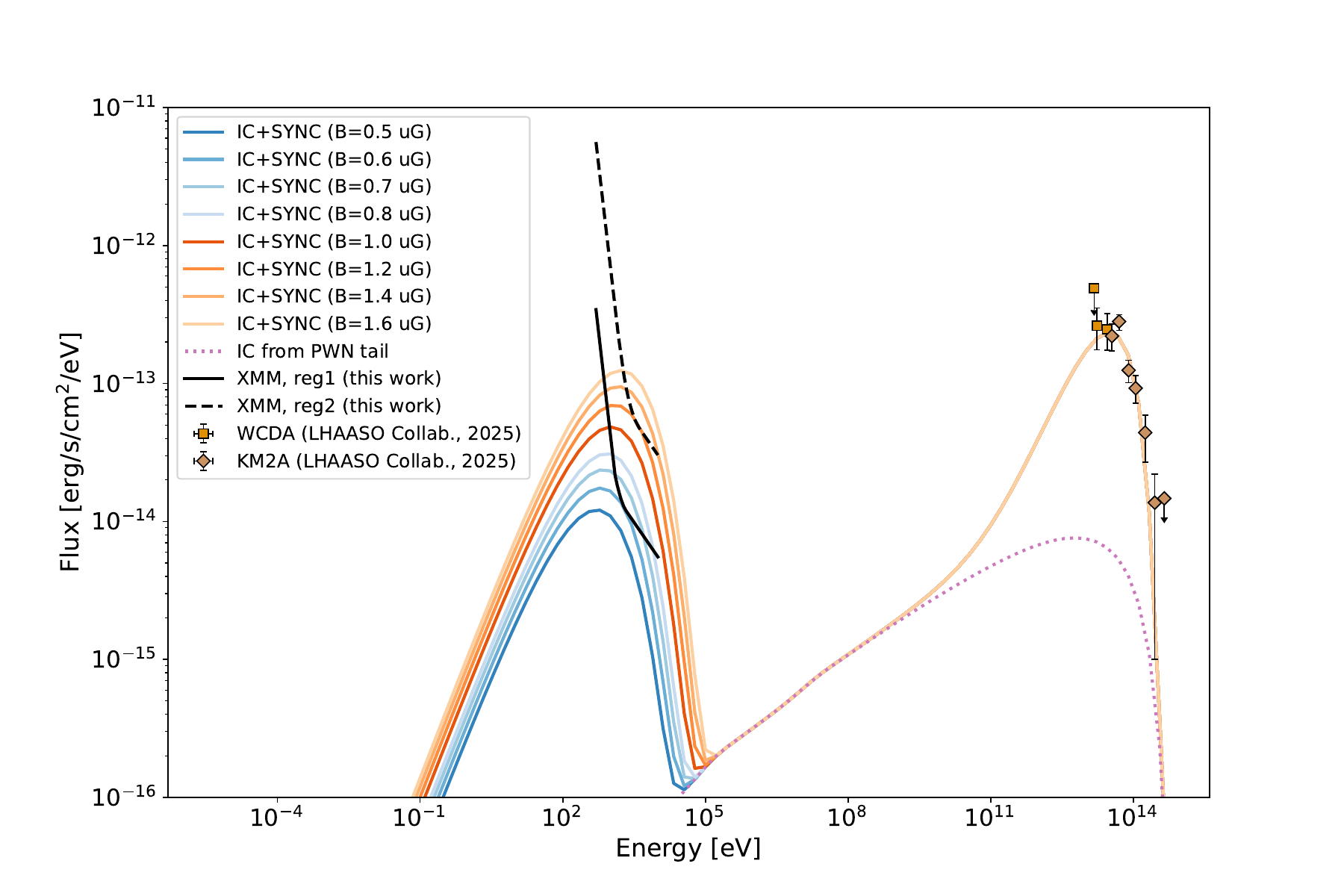}
    \includegraphics[width=0.49\linewidth]{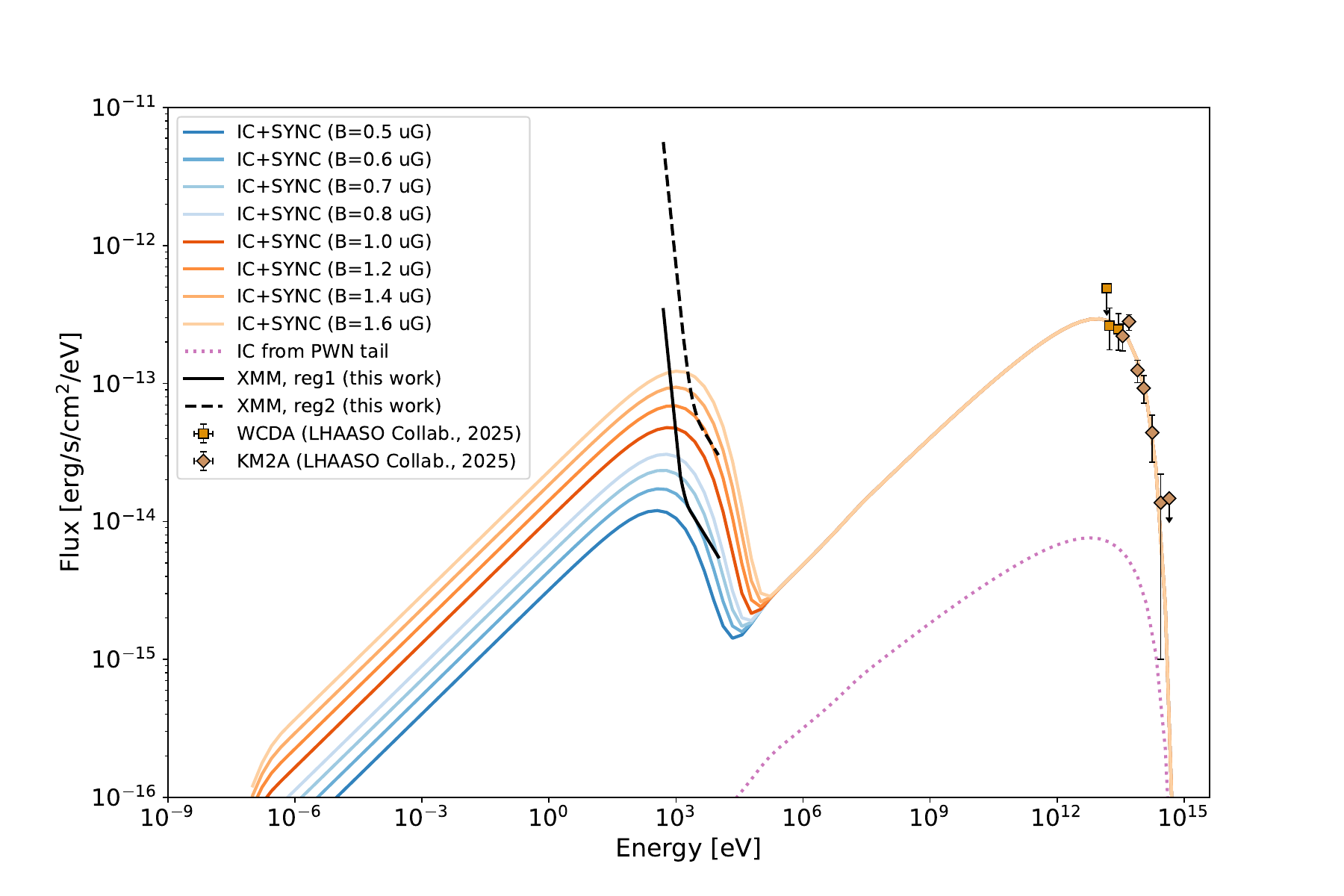}    
    \caption{SED model of the LHAASO spectrum along with the X-ray ULs for Region 1 (solid UL) and Region 2 (dashed UL) using an ECPL distribution for the relativistic electrons generating the emission with fixed spectral index $p$. The contribution to the ICS from the X-ray tail is shown as the dash-dotted line. Values of the magnetic field between 0.5 $\mu$G and 1.6 $\mu$G are tested. \textit{Left panel:} plot for $p=1.2$. \textit{Right panel:} plot for $p=2.4$. In both panels, the upper limits were computed by rescaling the results to the area of the LHAASO region, assuming a radius of 0.147$^\circ$ as \cite{LHAASO_2025}.}
    \label{fig:allB_plot}
\end{figure*}

The second hypothesis we investigated was whether 1LHAASO J1740+0948u could be associated with the diffuse X-ray emission in the surroundings of J1740. To evaluate this scenario, we assumed that both the LHAASO source and the diffuse X-ray emission originated from the same relativistic population of electrons and adopted the same ECPL distribution with fixed $E_0=1$ TeV, $\beta=2$, $E_{min}=1$ GeV and $E_{max}=960$ TeV as for the previous case. Since there is no constraint on the spectral index from the X-ray results, we first froze it and tested values between $p=1.2$ and $p=2.4$, then we let it free to vary. We also included in the fit the ICS emission from the X-ray tail, assuming the whole emission is due to the synchrotron-emitting particles, i.e. $p=2.52$, to take into account its possible contribution inside the LHAASO energy range. To model the synchrotron emission, we tested different values from 0.5 to 1.6 $\mu$G to determine the maximum value of the magnetic field of the diffuse emission before overshooting the X-ray limit. In Tab. \ref{tab:bestFit_diffuse} we report the best-fit values for $E_{cut}$ (in both electron and photon energy) and $W_e$ obtained with \texttt{naima} when fixing $p$ of the ECPL in the range 1.2--2.4, as well as the maximum logarithmic likelihood and the BIC value derived during the fit. We depicted in Figure \ref{fig:allB_plot} the models for $p=1.2$ (left) and $p=2.4$ (right). The derived values of both the BIC and maximum logarithmic likelihood slightly favour the models with the highest values of $p$. This is also confirmed by the best-fit result when using a free spectral index, for which we obtained $p=2.2 \pm 0.2$. Leaving $p$ free, however, did not improve the quality of the fit. In all the tested models, the values of the magnetic field required not to overshoot the X-ray upper limit are $\sim 0.6\ \mu$G for Region 1 and $\sim 1.1-1.2 \ \mu$G for Region 2, respectively. The different values in the two regions are likely due to the excess of non-thermal emission detected in Region 2 that causes the upper limits to be higher. Another possibility is related to the fact that Region 2 is located on the edge of the \textit{XMM-Newton} FoV, where the sensitivity is lower than at the centre, so the estimation could be affected by additional uncertainties. Such low values of the magnetic field are compatible with the hypothesis of them originating from an old electron population, too faint to be X-ray-emitting but only visible in the tera-electronvolt domain.

\begin{table}[ht]
    \centering
    \caption{Best-fit results for the SED fitting of the LHAASO spectrum and the \textit{XMM-Newton} upper limits on the diffuse emission when fixing the spectral index $p$.}
    \label{tab:bestFit_diffuse}
    \resizebox{\linewidth}{!}{%
    \begin{tabular}{cccccc}
        \toprule
        \toprule
         $p$ & $E_{cut}$ (TeV) & $E_{cut, \gamma}$ (TeV) & We (erg) & Max $\log ( \mathcal{L})$ & BIC \\
         \midrule
         1.2 & $156^{+15}_{-20}$ & $33 \pm 2$ & $1.87 \cdot 10^{44}$ & -3.46 & 11.53 \\
         1.4 & $164^{+14}_{-18}$ & $35^{+1}_{-2}$ & $2.29 \cdot 10^{44}$ & -3.29 & 11.19 \\
         1.6 & $163^{+28}_{-9}$ & $35.1^{+3.6}_{-0.8}$ & $3.29 \cdot 10^{44}$ & -3.09 & 10.78 \\
         1.8 & $196^{+26}_{-19}$ & $45^{+3}_{-9}$ & $4.75 \cdot 10^{44}$ & -2.93 & 10.46 \\
         2.0 & $199^{+26}_{-29}$ & $46^{+3}_{-4}$ & $1.20 \cdot 10^{45}$ & -2.78 & 10.16 \\
         2.2 & $231^{+30}_{-38}$ & $55^{+4}_{-5}$ & $4.16 \cdot 10^{45}$ & -2.68 & 9.96 \\
         2.4 & $247^{+39}_{-24}$ & $60^{+5}_{-3}$ & $2.24 \cdot 10^{46}$ & -2.65 & 9.90 \\
         \bottomrule
    \end{tabular}%
    }
    \tablefoot{In addition to the best-fit values for the cut-off energy $E_{cut}$, its correspondent photon energy $E_{cut, \gamma}$ and the total energy content $W_e$, we report the values of the maximum logarithmic likelihood and of the Bayesian information criterion (BIC) obtained with the \texttt{naima} fitting process.}
\end{table}

\subsection{Cooling times and distances}
To further support the hypothesis that the two populations, i.e. the one of the X-ray tail and that of the LHAASO source, are different, we estimated their cooling times. We followed the prescriptions of \cite{Khangulyan_2014} to compute the cooling times, and we used Eq. (1) of \cite{VanEtten_2011} to estimate the particle energy from the kilo-electronvolt-emitting photons and Eq. (1) of \cite{LHAASO_crab_2021} for the tera-electronvolt photons. We obtained a total cooling time for the particles observed in the energy range 0.5--8 keV of $\tau_{tot}\sim 1.2-4.3$ kyr, assuming the best-fit value of the PWN magnetic field obtained with \texttt{naima}. On the other hand, we derived a total cooling time in the range $\tau_{tot}\sim10-26$ kyr for the tera-electronvolt-emitting particles, assuming a magnetic field of 1 $\mu$G. If we focus on the synchrotron cooling time only, we obtained $\tau_{syn}\sim1.2-4.7$ kyr for the X-ray population and $\tau_{syn}\sim13-100$ kyr for the VHE population.

In Fig. \ref{fig:tails_with_dist} we see no spectral evolution with the distance from the pulsar, meaning that no cooling is detected along the entire visible length of the tail. To further support this finding, we computed the cooling distances of the electrons of the tail assuming different regimes for the particle transport, ballistic transport ($v=c$), advection ($v=c/\sqrt{3}$), and diffusion. For the latter, we used the following relation
\begin{equation}
    D(E)=D_0 \bigg( \frac{E}{10 \ \text{GeV}} \bigg)^{\frac{1}{3}},
\end{equation}
where $D_0=(3-8) \ \cdot 10^{28}$ cm$^2 \cdot$ s$^{-1}$, taken from \cite{Giacinti_2018}, and we chose $D_0=5\cdot 10^{28}$ cm$^2 \cdot$ s$^{-1}$ as the reference value at 10 GeV. We found distances in the range 0.4--1.3 kpc, 200--750 pc, and 88--140 pc for the ballistic transport, advection and diffusion, respectively. We also tested a slow-diffusion scenario, assuming $D_0=1.2\cdot 10^{26}$ cm\textsuperscript{2} $\cdot$ s\textsuperscript{-1} at 1 GeV, which is the average value of the best-fit results reported in \cite{Albert_2024} for the Geminga and Monogem halos. In this case, we obtained distances in the range of 6.1--9.6 pc. All the derived distances are larger than the physical offset observed for the LHAASO source, corresponding to $d = 4.7$ pc for a distance of J1740 of 1.23 kpc ($d=5.4$ pc for 1.4 kpc used in the \cite{LHAASO_2025} paper), suggesting particles could be able to travel to the LHAASO location without invoking any re-acceleration process.

\section{Discussion and conclusions} \label{sec:discussion}
In this work, we conducted a deep analysis of the X-ray emission around PSR J1740+1000 (J1740), a faint radio and gamma-ray pulsar located outside the Galactic plane, using $\sim500$ ks of \textit{XMM-Newton} observations. We re-analysed the long tail-like BSPWN and investigated for the first time the evolution of the spectral index as a function of the distance from J1740. We also performed the first study of the diffuse emission in the surroundings of J1740, using a custom model for the background emission. Our goal was to constrain a possible X-ray association with the recently detected UHE source 1LHAASO J1740+0948u, located at a 0.22$^\circ$ offset from the pulsar position.

The analysis of the cometary tail proved the best-fit model in the energy range 0.5--8 keV to be an absorbed power law with photon index $\Gamma=1.76 \pm 0.06$, compatible with the result obtained by \cite{Benbow_2021} on the same dataset but with an independent analysis. The efficiency of the PWN, derived from the X-ray luminosity and assuming a dispersion measure distance of 1.23 kpc, is $\eta_X=1.5 \cdot 10^{-4}$, a typical value observed for middle-aged pulsars \citep{Kargaltsev_Pavlov_2008}. For the first time, we explored the evolution of the spectral index with distance and found no evidence for spectral cooling along the length of the tail. A decrease in the absolute value of the spectral index is found in the outermost region of the tail, at an angular distance of $\sim0.1^\circ$ from the pulsar position, but the poor statistic implies large error bars on the result and makes the result still compatible with that of the other regions. It is still not well understood why some BSPWNe, like the Lighthouse PWN \citep{Pavan_2016} or the Mouse PWN \citep{Klingler_2018}, show a spectral evolution, while other systems, such as the one of J1740 studied in this work or that of PSR B0355+54 \citep{Klingler_2016}, do not show this feature. A possible explanation could be related to the different environments where the PWNe are located or different interactions between the particles and the surrounding medium, possibly connected to the age of the population.

For the study of the diffuse emission, instead, we focused on two regions of the \textit{XMM-Newton} FoV, due to the limited coverage of the extension of 1LHAASO J1740+0948u provided by the instrument. The first (Region 1) is a $r = 0.11^\circ$ circular region centred on the pulsar and excluding both the tail and the pulsar to avoid their contamination. The second (Region 2) is a $r\sim0.06^\circ$ circle comprising the majority of the 68\% extension of the VHE source. We fitted custom models for the instrumental and the physical backgrounds, where the latter comprised several components: the cosmic X-ray background, the Galactic halo, the local hot bubble and the North Polar Spur, a region of enhanced X-ray emission whose position coincides with J1740's location. To reproduce the non-thermal diffuse component we aimed to study, we added an absorbed power law corresponding to the source term. No detection was found for Region 1, while a weak component was detected for Region 2. This component, however, was only marginally detected, and deeper observations or coverage of a more extended area would be required to firmly exclude that such a feature is not due to a background fluctuation. Therefore, we treated it as a non-detection. We then derived the 3$\sigma$ upper limits on the emission in the range 0.5--10 keV for both regions by combining the results obtained for fixed values of the photon index between $\Gamma=0.5$ and $\Gamma=3.0$. To constrain the nature of 1LHAASO J1740+0948u, we investigated two main hypotheses: (1) LHAASO detected the X-ray tail in the VHE band, and (2) the VHE source may be related to the diffuse non-thermal X-ray emission.

To explore hypothesis (1), we performed an SED fitting of the tail emission together with the spectrum displayed in \cite{LHAASO_2025}, assuming the particles to be distributed as an exponential cut-off power law. The approach is similar to that presented in \cite{Benbow_2021}, published before the discovery of 1LHAASO J1740+0948u, but we were able to further improve it. We performed a multi-wavelength MCMC fitting, including both a spectrum in the X-rays and in the VHEs, allowing for a better constraint of the parameters of the electron distribution when compared to a fit of the X-ray data alone. The model obtained when fixing the spectral index to the expected value from the X-ray emission, i.e. $p=2.52$, is not able to fit the VHE part of the SED, but constrains the tail magnetic field to $B=6.8 \pm 1.9 \ \mu$G, compatible with the predictions by models of BSPWNe \citep{Olmi_Bucciantini_2019} and with previous literature results on J1740 \citep{Benbow_2021, Kargaltsev_2008}. The projected location of J1740 in the sky is coincident with the NPS (see Fig. \ref{fig:fov_xmm}), and what we obtained for the tail magnetic field is of the same order as the average value measured for the NPS \citep{Iwashita_2023}. However, the distance to the pulsar as measured by \cite{Yao_2017} is $d\sim 1.23$ kpc, whereas the root of NPS is measured to have a distance of several kilo-parsecs \citep{Zhang_2024}. Therefore, the magnetic field measured from SED fitting in this work corresponds to the vicinity of the PWN, not the NPS. In order to have a better fit of the LHAASO spectrum, we conducted several tests and found that the model with $p=1.6$ returned the best-fit results, both in the match with the VHE points and in the results of the BIC and likelihood. This index is considerably harder than the expected $p=2.52$ and underestimates both the magnetic field of the tail and the VHE flux at $\sim10$ TeV by a factor of two or three. These results show that LHAASO likely did not detect the counterpart of the X-ray tail and that the two populations, the X-ray bright one and that observed in the VHEs, are distinct: the VHE emission traces older electrons that are not emitting in the X-rays any more due to cooling, while the tail is populated by young and fresh electrons injected from the pulsar.

We then explored hypothesis (2) by carrying out an SED fitting using the spectrum of 1LHAASO J1740+0948u \cite{LHAASO_2025} and the X-ray upper limits on the diffuse emission obtained for Regions 1 and 2. With this approach, it was possible to constrain the magnetic field to $B\leq0.6 \ \mu$G for Region 1 and to $B\leq 1.1-1.2 \ \mu$G for Region 2, where the difference between the second and the first value could be due to the non-thermal excess marginally detected for Region 2, to physical fluctuations of the magnetic field in the medium or to the result being affected by the worse performance of \textit{XMM-Newton} when observing off-axis. Similar methods for the SED modelling were adopted in \cite{LHAASO_2025} and \cite{Xie_2025}. Both works attempted to put constraints on the electron population generating the LHAASO emission and its possible X-ray counterparts. In our study, we included in the picture the very first upper limits on the diffuse emission as obtained by \textit{XMM-Newton} and we constrained the magnetic field of the diffuse emission based on the real observations. In addition, we relaxed some of the assumptions that were used in both previous works for the SED modelling parameters.

Considering these results, we argue that the X-ray tail is produced via synchrotron radiation of freshly injected electrons in a magnetic field of $\sim6.8 \ \mu$G, while the LHAASO emission traces older electrons. This population is not emitting in the X-rays anymore due to cooling and the decrease of the magnetic field to $\leq 0.6-1.2\ \mu$G, but it is still shining via ICS of the ambient photon fields. One possible interpretation of the nature of this system is that the LHAASO emission represents the relic PWN of J1740. For these sources, it is common to observe offsets between the centroid of the VHE source and the pulsar position or the X-ray nebula, if detected. This is compatible with the 0.22$^\circ$ displacement observed between 1LHAASO J1740+0948u and the radio and X-ray coordinates of J1740. There are several examples of relic PWNe systems of different ages showing similar offsets: HESS J1825--137, showing an extended asymmetric morphology 0.2$^\circ$ offset from PSR J1826--1334 and its X-ray nebula \citep{HESSJ1825_2006, HESSJ1825_2019}; HESS J1616--508, without a detected counterpart in the X-rays and located at $\sim10'$ from PSR J1617--5055 and its faint PWN \citep{Kargaltsev_2009}; HESS J1356--645, associated with PSR J1357--6429, which hosts an extended nebula and is located at a distance of $\sim7'$ from the VHE source \citep{HESS_J1356}; or HESS J1303--631, which centroid is at $\sim3.1'$ from PSR J1301--6305 and was one of the first discovered `dark' sources without an X-ray counterpart \citep{HESS_J1303}. There are two main explanations for the observed offsets between X-rays and VHEs. They could be either caused by asymmetries in the medium and, consequently, in the SNR reverse shock that interacts with the expanding PWN \citep{Blondin_2001}, or due to the supersonic motion of the pulsar related to the kick velocity that it has acquired during its birth \citep{Fiori_2022}. Considering J1740 is classified as a BSPWN, the second scenario seems to be the most likely, even though an asymmetric expansion of the parent SNR, not detected either in radio or X-rays so far, in principle could have occurred in the past. Considering that the $\sim0.22^\circ$ offset between the LHAASO centroid and the position of J1740 corresponds to only $\sim5$ pc, for both $d=1.23$ kpc used in this paper and $d=1.4$ kpc adopted by \cite{LHAASO_2025}, this kind of interpretation does not need to invoke any re-acceleration process to explain the connection between the X-ray upper limit and the VHE emission.

The other possible scenario is that 1LHAASO J1740+0948u represents the pulsar halo of J1740. This could be a realistic hypothesis considering the age ($\tau_c=114$ kyr) and its classification as a BSPWN that has likely already escaped the parent SNR and is interacting directly with the ISM. It is interesting to notice that the High Altitude Water Cherenkov (HAWC) experiment detected a weak (TS=28) VHE source near J1740, dubbed 3HWC J1739+099, but it is a point-like source and its centre is off-site compared to the LHAASO source \citep{Albert_2020}. The source, which has been classified as a candidate pulsar halo, has not been included in our work due to the uncertainty in the connection with 1LHAASO J1740+0948u and the lack of spectral points of the source. To further investigate the interpretation of 1LHAASO J1740+0948u as a pulsar halo, we followed a phenomenological approach. A criterion to understand whether a VHE source can be classified as a pulsar halo is to compare its energy density to that of the ISM: if the former is lower than the latter, it is evidence for outflowing electrons towards a region not influenced by the pulsar \citep{LopezCoto_2022}. We computed the energy density of 1LHAASO J1740+0948u by following the method explained in \cite{Giacinti_2020}, using the $W_e$ value computed by \texttt{naima} instead of performing the integral of the spectrum. We assumed $r=0.147^\circ$, which is the 95\% confidence level upper limit on the source size under the point-like source assumption \citep{LHAASO_2025}, as the radius of the region. The results for the spectral indices between $p=1.2$ and $p=2.4$ are in the range $0.03 - 3.6$ eV/cm\textsuperscript{3}. Excluding the extreme value obtained for $p=2.4$, which is likely a too soft index for an injection spectrum, the energy densities are in the range $0.03 - 0.67$ eV/cm\textsuperscript{3}. When leaving the spectral index free to vary in the fit, we obtained $p=2.2 \pm 0.4$ with a corresponding $E_e=3.45 \cdot 10^{45}$ erg and an energy density of $\varepsilon=0.57$ eV/cm\textsuperscript{3}. The conventional value assumed for the ISM energy density is 0.1 eV/cm\textsuperscript{3}, but sources with energy densities lower than $\sim$ 1 eV/cm\textsuperscript{3} may already display halo-like features. When studying the cooling of the tail electrons, we found they can reach very large distances from the pulsar. Assuming slow diffusion, instead, we derived a range of $\sim 6 - 9$ pc, which is larger than the offset than the physical offset observed between the pulsar and the LHAASO centroid ($d\sim 5$ pc). Even in a slow diffusion context, electrons may be able to reach the LHAASO location and generate a pulsar halo. Moreover, projection effects could explain both the offset and the small size, as already discussed in \cite{LHAASO_2025}. Therefore, we argue 1LHAASO J1740+0948u may likely be the pulsar halo of J1740 and, if this hypothesis is confirmed, we constrained the magnetic field around the pulsar to be as low as $B\leq 0.6 - 1.2\ \mu$G, lower than the typical value of the ISM magnetic field and compatible with other results obtained for the search of X-ray pulsar halos \citep{Khokhriakova_2024, Manconi_2024, Adams_2025}. Future on-site X-ray observations should help unravel the nature of this UHE source.

Due to the lack of multi-wavelength coverage on the source and the absence of dedicated X-ray observations covering the full extension of 1LHAASO J1740+0948u, it was beyond the aims of this paper to develop a detailed treatment of diffusion to further support the pulsar halo scenario. A more complicated modelling, combining a spectral and spatial analysis of the multi-wavelength emission, is likely needed to investigate the source more in detail and have a better understanding of the relation between the X-ray bow-shock tail and the LHAASO emission. Future observations by particle detectors, such as LHAASO and HAWC, or by the upcoming Cherenkov Telescope Array Observatory, could help us understand more about this dark VHE source and whether the two detections, i.e. 1LHAASO J1740+0948u and 3HWC J1739+099, represent the same source. Future X-ray facilities, such as \textit{NewAthena} \citep{Cruise_2025}, could allow us to distinguish better the diffuse emission around J1740 from the background and improve the upper limits obtained in this work.

\begin{acknowledgements}    
    This research is based on observations performed by the satellite \textit{XMM-Newton}, a European Space Agency science mission with instruments and contributions directly funded by ESA Member States and NASA. This work has also made use of data and software provided by the High Energy Astrophysics Science Archive Research Center (HEASARC), which is a service of the Astrophysics Science Division at NASA/GSFC. GP and HZ acknowledge financial support from the European Research Council (ERC) under the European Union’s Horizon 2020 research and innovation program HotMilk (grant agreement No. 865637). GP also acknowledges support from Bando per il Finanziamento della Ricerca Fondamentale 2022 dell’Istituto Nazionale di Astrofisica (INAF): GO Large program and from the Framework per l’Attrazione e il Rafforzamento delle Eccellenze (FARE) per la ricerca in Italia (R20L5S39T9).
\end{acknowledgements}

%
%

\bibliographystyle{aa} 
\bibliography{biblio} 

\begin{thebibliography}{70}
\expandafter\ifx\csname natexlab\endcsname\relax\def\natexlab#1{#1}\fi

\bibitem[{Abdollahi {et~al.}(2022)Abdollahi, Acero, Baldini, Ballet, Bastieri, Bellazzini, Berenji, Berretta, Bissaldi, Blandford, Bloom, Bonino, Brill, Britto, Bruel, Burnett, Buson, Cameron, Caputo, Caraveo, Castro, Chaty, Cheung, Chiaro, Cibrario, Ciprini, Coronado-Blázquez, Crnogorcevic, Cutini, D’Ammando, Gaetano, Digel, Lalla, Dirirsa, Venere, Domínguez, Ramazani, Fegan, Ferrara, Fiori, Fleischhack, Franckowiak, Fukazawa, Funk, Fusco, Galanti, Gammaldi, Gargano, Garrappa, Gasparrini, Giacchino, Giglietto, Giordano, Giroletti, Glanzman, Green, Grenier, Grondin, Guillemot, Guiriec, Gustafsson, Harding, Hays, Hewitt, Horan, Hou, Jóhannesson, Karwin, Kayanoki, Kerr, Kuss, Landriu, Larsson, Latronico, Lemoine-Goumard, Li, Liodakis, Longo, Loparco, Lott, Lubrano, Maldera, Malyshev, Manfreda, Martí-Devesa, Mazziotta, Mereu, Meyer, Michelson, Mirabal, Mitthumsiri, Mizuno, Moiseev, Monzani, Morselli, Moskalenko, Negro, Nuss, Omodei, Orienti, Orlando, Paneque, Pei, Perkins, Persic, Pesce-Rollins, Petrosian,
  Pillera, Poon, Porter, Principe, Rainò, Rando, Rani, Razzano, Razzaque, Reimer, Reimer, Reposeur, Sánchez-Conde, Parkinson, Scotton, Serini, Sgrò, Siskind, Smith, Spandre, Spinelli, Sueoka, Suson, Tajima, Tak, Thayer, Thompson, Torres, Troja, Valverde, Wood, \& Zaharijas}]{Abdollahi_2022}
Abdollahi, S., Acero, F., Baldini, L., {et~al.} 2022, \apjs, 260, 53

\bibitem[{Abeysekara {et~al.}(2017)Abeysekara, Albert, Alfaro, Alvarez, Álvarez, Arceo, Arteaga-Velázquez, Rojas, Solares, Barber, Bautista-Elivar, Becerril, Belmont-Moreno, BenZvi, Berley, Bernal, Braun, Brisbois, Caballero-Mora, Capistrán, Carramiñana, Casanova, Castillo, Cotti, Cotzomi, de~León, León, la~Fuente, Dingus, DuVernois, Díaz-Vélez, Ellsworth, Engel, Enríquez-Rivera, Fiorino, Fraija, García-González, Garfias, Gerhardt, Muñoz, González, Goodman, Hampel-Arias, Harding, Hernández, Hernández-Almada, Hinton, Hona, Hui, Hüntemeyer, Iriarte, Jardin-Blicq, Joshi, Kaufmann, Kieda, Lara, Lauer, Lee, Lennarz, Vargas, Linnemann, Longinotti, Raya, Luna-García, López-Coto, Malone, Marinelli, Martinez, Martinez-Castellanos, Martínez-Castro, Martínez-Huerta, Matthews, Miranda-Romagnoli, Moreno, Mostafá, Nellen, Newbold, Nisa, Noriega-Papaqui, Pelayo, Pretz, Pérez-Pérez, Ren, Rho, Rivière, Rosa-González, Rosenberg, Ruiz-Velasco, Salazar, Greus, Sandoval, Schneider, Schoorlemmer, Sinnis,
  Smith, Springer, Surajbali, Taboada, Tibolla, Tollefson, Torres, Ukwatta, Vianello, Weisgarber, Westerhoff, Wisher, Wood, Yapici, Yodh, Younk, Zepeda, Zhou, Guo, Hahn, Li, \& Zhang}]{HAWC_2017}
Abeysekara, A.~U., Albert, A., Alfaro, R., {et~al.} 2017, Science, 358, 911

\bibitem[{Adams {et~al.}(2025)Adams, Archer, Bangale, Bartkoske, Benbow, Buckley, Chen, Christiansen, Chromey, Duerr, Errando, Escobar~Godoy, Falcone, Feldman, Feng, Fortson, Furniss, Hanlon, Hervet, Hinrichs, Holder, Humensky, Jin, Johnson, Kaaret, Kertzman, Kherlakian, Kieda, Kleiner, Korzoun, Krennrich, Kumar, Kundu, Lang, Lundy, Maier, Millard, Millis, Mooney, Moriarty, Mukherjee, Ning, Ong, Pandey, Pohl, Pueschel, Quinn, Rabinowitz, Ragan, Reynolds, Ribeiro, Rizk, Roache, Sadeh, Saha, Sembroski, Shang, Splettstoesser, Tak, Talluri, Tucci, Valverde, Williams, Wong, Woo, collaboration), Kwong, Mori, Hailey, Safi-Harb, Zhang, Tsuji, collaboration), Manconi, Donato, \& Di~Mauro}]{Adams_2025}
Adams, C.~B., Archer, A., Bangale, P., {et~al.} 2025, \apj, 985, 90

\bibitem[{Aharonian {et~al.}(2006)Aharonian, Akhperjanian, Bazer-Bachi, {Beilicke, M.}, {Benbow, W.}, {Berge, D.}, {Bernlöhr, K.}, {Boisson, C.}, {Bolz, O.}, {Borrel, V.}, {Braun, I.}, {Brown, A. M.}, {Bühler, R.}, {Büsching, I.}, {Carrigan, S.}, {Chadwick, P. M.}, {Chounet, L.-M.}, {Cornils, R.}, {Costamante, L.}, {Degrange, B.}, {Dickinson, H. J.}, {Djannati-Ataï, A.}, {Drury, L. O'C.}, {Dubus, G.}, {Egberts, K.}, {Emmanoulopoulos, D.}, {Espigat, P.}, {Feinstein, F.}, {Ferrero, E.}, {Fiasson, A.}, {Fontaine, G.}, {Funk, Seb.}, {Funk, S.}, {Füßling, M.}, {Gallant, Y. A.}, {Giebels, B.}, {Glicenstein, J. F.}, {Goret, P.}, {Hadjichristidis, C.}, {Hauser, D.}, {Hauser, M.}, {Heinzelmann, G.}, {Henri, G.}, {Hermann, G.}, {Hinton, J. A.}, {Hoffmann, A.}, {Hofmann, W.}, {Holleran, M.}, {Horns, D.}, {Jacholkowska, A.}, {de Jager, O. C.}, {Kendziorra, E.}, {Khélifi, B.}, {Komin, Nu.}, {Konopelko, A.}, {Kosack, K.}, {Latham, I. J.}, {Le Gallou, R.}, {Lemière, A.}, {Lemoine-Goumard, M.}, {Lohse, T.}, {Martin,
  J. M.}, {Martineau-Huynh, O.}, {Marcowith, A.}, {Masterson, C.}, {Maurin, G.}, {McComb, T. J. L.}, {Moulin, E.}, {de Naurois, M.}, {Nedbal, D.}, {Nolan, S. J.}, {Noutsos, A.}, {Orford, K. J.}, {Osborne, J. L.}, {Ouchrif, M.}, {Panter, M.}, {Pelletier, G.}, {Pita, S.}, {Pühlhofer, G.}, {Punch, M.}, {Raubenheimer, B. C.}, {Raue, M.}, {Rayner, S. M.}, {Reimer, A.}, {Reimer, O.}, {Ripken, J.}, {Rob, L.}, {Rolland, L.}, {Rowell, G.}, {Sahakian, V.}, {Santangelo, A.}, {Saugé, L.}, {Schlenker, S.}, {Schlickeiser, R.}, {Schröder, R.}, {Schwanke, U.}, {Schwarzburg, S.}, {Shalchi, A.}, {Sol, H.}, {Spangler, D.}, {Spanier, F.}, {Steenkamp, R.}, {Stegmann, C.}, {Superina, G.}, {Tavernet, J.-P.}, {Terrier, R.}, {Théoret, C. G.}, {Tluczykont, M.}, {van Eldik, C.}, {Vasileiadis, G.}, {Venter, C.}, {Vincent, P.}, {Völk, H. J.}, {Wagner, S. J.}, \& {Ward, M.}}]{HESSJ1825_2006}
Aharonian, F., Akhperjanian, A.~G., Bazer-Bachi, A.~R., {et~al.} 2006, A\&A, 460, 365

\bibitem[{Aharonian {et~al.}(2021)Aharonian, An, Axikegu, Bai, Bai, Bao, Bastieri, Bi, Bi, Cai, Cai, Cao, Cao, Chang, Chang, Chang, Chen, Chen, Chen, Chen, Chen, Chen, Chen, Chen, Chen, Chen, Chen, Chen, Chen, Cheng, Cheng, Cui, Cui, Cui, Dai, Dai, Dai, Danzengluobu, della Volpe, D'Ettorre~Piazzoli, Dong, Fan, Fan, Fan, Fang, Fang, Feng, Feng, Feng, Feng, Gao, Gao, Gao, Gao, Ge, Geng, Gong, Gou, Gu, Guo, Guo, Guo, Guo, Han, He, He, He, He, He, He, Heller, Hor, Hou, Hou, Hu, Hu, Hu, Hu, Huang, Huang, Huang, Huang, Huang, Ji, Ji, Jia, Jiang, Jiang, Jin, Kuleshov, Levochkin, Li, Li, Li, Li, Li, Li, Li, Li, Li, Li, Li, Li, Li, Li, Li, Li, Li, Liang, Liang, Lin, Liu, Liu, Liu, Liu, Liu, Liu, Liu, Liu, Liu, Liu, Liu, Liu, Liu, Liu, Liu, Long, Lu, Lv, Ma, Ma, Ma, Mao, Masood, Mitthumsiri, Montaruli, Nan, Pang, Pattarakijwanich, Pei, Qi, Ruffolo, Rulev, S\'aiz, Shao, Shchegolev, Sheng, Shi, Song, Stenkin, Stepanov, Sun, Sun, Sun, Tam, Tang, Tian, Wang, Wang, Wang, Wang, Wang, Wang, Wang, Wang, Wang, Wang, Wang, Wang,
  Wang, Wang, Wang, Wang, Wang, Wang, Wang, Wang, Wang, Wei, Wei, Wei, Wen, Wu, Wu, Wu, Wu, Wu, Xi, Xia, Xia, Xiang, Xiao, Xiao, Xin, Xin, Xing, Xu, Xu, Xue, Yan, Yang, Yang, Yang, Yang, Yang, Yang, Yang, Yao, Yao, Ye, Yin, Yin, You, You, Yu, Yuan, Zeng, Zeng, Zeng, Zeng, Zha, Zhai, Zhang, Zhang, Zhang, Zhang, Zhang, Zhang, Zhang, Zhang, Zhang, Zhang, Zhang, Zhang, Zhang, Zhang, Zhang, Zhang, Zhang, Zhang, Zhang, Zhao, Zhao, Zhao, Zhao, Zhao, Zheng, Zheng, Zhou, Zhou, Zhou, Zhou, Zhou, Zhou, Zhu, Zhu, Zhu, Zhu, Zuo, \& Huang}]{LHAASO_2021}
Aharonian, F., An, Q., Axikegu, {et~al.} 2021, Phys. Rev. Lett., 126, 241103

\bibitem[{Albert {et~al.}(2024)Albert, Alfaro, Alvarez, Arteaga-Velázquez, Avila~Rojas, Ayala~Solares, Babu, Belmont-Moreno, Bernal, Caballero-Mora, Capistrán, Carramiñana, Casanova, Cotti, Cotzomi, Coutiño~de León, de~la Fuente, Depaoli, Di~Lalla, Diaz~Hernandez, Dingus, DuVernois, Durocher, Díaz-Vélez, Engel, Espinoza, Fan, Fang, Fraija, García-González, Garfias, Goksu, González, Goodman, Groetsch, Harding, Hernández-Cadena, Herzog, Hüntemeyer, Huang, Hueyotl-Zahuantitla, Iriarte, Joshi, Kaufmann, Kieda, Lara, Lee, Lee, León~Vargas, Linnemann, Longinotti, Luis-Raya, Malone, Martinez, Martínez-Castro, Matthews, Miranda-Romagnoli, Montes, Morales-Soto, Moreno, Mostafá, Nayerhoda, Nellen, Noriega-Papaqui, Olivera-Nieto, Omodei, Pérez~Araujo, Pérez-Pérez, Rho, Rosa-González, Salazar, Salazar-Gallegos, Sandoval, Schneider, Schwefer, Serna-Franco, Son, Springer, Tibolla, Tollefson, Torres, Torres-Escobedo, Turner, Urea-Mena, Varela, Villaseñor, Wang, Watson, Willox, Wu, Yun-Cárcamo, Zhou,
  de~León, (for~theHAWC Collaboration), \& Di~Mauro}]{Albert_2024}
Albert, A., Alfaro, R., Alvarez, C., {et~al.} 2024, \apj, 974, 246

\bibitem[{Albert {et~al.}(2020)Albert, Alfaro, Alvarez, Camacho, Arteaga-Velázquez, Arunbabu, Rojas, Solares, Baghmanyan, Belmont-Moreno, BenZvi, Brisbois, Caballero-Mora, Capistrán, Carramiñana, Casanova, Cotti, de~León, la~Fuente, Hernandez, Diaz-Cruz, Dingus, DuVernois, Durocher, Díaz-Vélez, Ellsworth, Engel, Espinoza, Fan, Fang, Alonso, Fleischhack, Fraija, Galván-Gámez, Garcia, García-González, Garfias, Giacinti, González, Goodman, Harding, Hernandez, Hinton, Hona, Huang, Hueyotl-Zahuantitla, Hüntemeyer, Iriarte, Jardin-Blicq, Joshi, Kieda, Lara, Lee, Vargas, Linnemann, Longinotti, Luis-Raya, Lundeen, López-Coto, Malone, Marandon, Martinez, Martinez-Castellanos, Martínez-Castro, Matthews, Miranda-Romagnoli, Morales-Soto, Moreno, Mostafá, Nayerhoda, Nellen, Newbold, Nisa, Noriega-Papaqui, Olivera-Nieto, Omodei, Peisker, Araujo, Pérez-Pérez, Ren, Rho, Rivière, Rosa-González, Ruiz-Velasco, Salazar, Greus, Sandoval, Schneider, Schoorlemmer, Serna, Sinnis, Smith, Springer, Surajbali,
  Tollefson, Torres, Torres-Escobedo, Ukwatta, Ureña-Mena, Weisgarber, Werner, Willox, Zepeda, Zhou, de~León, Álvarez, \& Collaboration)}]{Albert_2020}
Albert, A., Alfaro, R., Alvarez, C., {et~al.} 2020, \apj, 905, 76

\bibitem[{Arnaud(1996)}]{Arnaud_1996}
Arnaud, K.~A. 1996, in Astronomical Society of the Pacific Conference Series, Vol. 101, Astronomical Data Analysis Software and Systems V, ed. G.~H. {Jacoby} \& J.~{Barnes}, 17

\bibitem[{Benbow {et~al.}(2021)Benbow, Brill, Buckley, Capasso, Chromey, Errando, Falcone, Farrell, Feng, Finley, Foote, Fortson, Furniss, Gent, Giuri, Hanna, Hassan, Hervet, Holder, Hughes, Humensky, Jin, Kaaret, Kargaltsev, Kertzman, Kieda, Klingler, Kumar, Lang, Lundy, Maier, McGrath, Moriarty, Mukherjee, Nieto, Nievas-Rosillo, O’Brien, Ong, Otte, Patel, Pfrang, Pohl, Prado, Quinn, Ragan, Reynolds, Ribeiro, Richards, Roache, Ryan, Santander, Sembroski, Shang, Volkov, Wakely, Weinstein, Wilcox, \& Williams}]{Benbow_2021}
Benbow, W., Brill, A., Buckley, J.~H., {et~al.} 2021, \apj, 916, 117

\bibitem[{Blondin {et~al.}(2001)Blondin, Chevalier, \& Frierson}]{Blondin_2001}
Blondin, J.~M., Chevalier, R.~A., \& Frierson, D.~M. 2001, \apj, 563, 806

\bibitem[{Breuhaus {et~al.}(2022)Breuhaus, Reville, \& Hinton}]{Breuhaus_2022}
Breuhaus, M., Reville, B., \& Hinton, J.~A. 2022, A\&A, 660, A8

\bibitem[{Cao {et~al.}(2024)Cao, Aharonian, An, Axikegu, Bai, Bao, Bastieri, Bi, Bi, Cai, Cao, Cao, Cao, Chang, Chang, Chen, Chen, Chen, Chen, Chen, Chen, Chen, Chen, Chen, Chen, Chen, Chen, Cheng, Cheng, Cui, Cui, Cui, Cui, Dai, Dai, Dai, Danzengluobu, della Volpe, Dong, Duan, Fan, Fan, Fang, Fang, Feng, Feng, Feng, Feng, Feng, Gabici, Gao, Gao, Gao, Gao, Gao, Gao, Ge, Geng, Giacinti, Gong, Gou, Gu, Guo, Guo, Guo, Guo, Han, He, He, He, He, He, Heller, Hor, Hou, Hou, Hou, Hu, Hu, Hu, Huang, Huang, Huang, Huang, Huang, Huang, Huang, Ji, Jia, Jia, Jiang, Jiang, Jiang, Jin, Kang, Ke, Kuleshov, Kurinov, Li, Li, Li, Li, Li, Li, Li, Li, Li, Li, Li, Li, Li, Li, Li, Li, Li, Li, Li, Liang, Liang, Lin, Liu, Liu, Liu, Liu, Liu, Liu, Liu, Liu, Liu, Liu, Liu, Liu, Liu, Liu, Lu, Luo, Lv, Ma, Ma, Ma, Mao, Min, Mitthumsiri, Mu, Nan, Neronov, Ou, Pang, Pattarakijwanich, Pei, Qi, Qi, Qiao, Qin, Ruffolo, Sáiz, Semikoz, Shao, Shao, Shchegolev, Sheng, Shu, Song, Stenkin, Stepanov, Su, Sun, Sun, Sun, Tam, Tang, Tang, Tian, Wang,
  Wang, Wang, Wang, Wang, Wang, Wang, Wang, Wang, Wang, Wang, Wang, Wang, Wang, Wang, Wang, Wang, Wang, Wang, Wang, Wang, Wei, Wei, Wei, Wen, Wu, Wu, Wu, Wu, Wu, Xi, Xia, Xia, Xiang, Xiao, Xiao, Xin, Xin, Xing, Xiong, Xu, Xu, Xu, Xu, Xue, Yan, Yan, Yan, Yang, Yang, Yang, Yang, Yang, Yang, Yang, Yang, Yang, Yao, Yao, Ye, Yin, Yin, You, You, Yu, Yuan, Yue, Zeng, Zeng, Zeng, Zha, Zhang, Zhang, Zhang, Zhang, Zhang, Zhang, Zhang, Zhang, Zhang, Zhang, Zhang, Zhang, Zhang, Zhang, Zhang, Zhang, Zhang, Zhang, Zhao, Zhao, Zhao, Zhao, Zhao, Zheng, Zhou, Zhou, Zhou, Zhou, Zhou, Zhou, Zhou, Zhu, Zhu, Zhu, Zhu, Zuo, \& Collaboration)}]{Cao_2024}
Cao, Z., Aharonian, F., An, Q., {et~al.} 2024, \apjs, 271, 25

\bibitem[{Cruise {et~al.}(2025)Cruise, Guainazzi, Aird, Carrera, Costantini, Corrales, Dauser, Eckert, Gastaldello, Matsumoto, Osten, Petrucci, Porquet, Pratt, Rea, Reiprich, Simionescu, Spiga, \& Troja}]{Cruise_2025}
Cruise, M., Guainazzi, M., Aird, J., {et~al.} 2025, Nature Astronomy, 9, 36

\bibitem[{de~Jager \& Djannati-Ata{\"i}(2009)}]{deJager_2009}
de~Jager, O.~C. \& Djannati-Ata{\"i}, A. 2009, Implications of HESS Observations of Pulsar (Berlin, Heidelberg: Springer Berlin Heidelberg), 451--479

\bibitem[{de~Oña~Wilhelmi {et~al.}(2022)de~Oña~Wilhelmi, López-Coto, Amato, \& Aharonian}]{deOnaWilhelmi_2022}
de~Oña~Wilhelmi, E., López-Coto, R., Amato, E., \& Aharonian, F. 2022, \apjl, 930, L2

\bibitem[{{ESA: XMM-Newton SOC}(2024)}]{xmm_handbook}
{ESA: XMM-Newton SOC}. 2024, XMM-Newton Users Handbook - Issue 2.22

\bibitem[{Fiori {et~al.}(2022)Fiori, Olmi, Amato, Bandiera, Bucciantini, Zampieri, \& Burtovoi}]{Fiori_2022}
Fiori, M., Olmi, B., Amato, E., {et~al.} 2022, \mnras, 511, 1439

\bibitem[{Giacinti {et~al.}(2018)Giacinti, Kachelrieẞ, \& Semikoz}]{Giacinti_2018}
Giacinti, G., Kachelrieẞ, M., \& Semikoz, D. 2018, Journal of Cosmology and Astroparticle Physics, 2018, 051

\bibitem[{Giacinti {et~al.}(2020)Giacinti, Mitchell, López-Coto, Joshi, Parsons, \& Hinton}]{Giacinti_2020}
Giacinti, G., Mitchell, A. M.~W., López-Coto, R., {et~al.} 2020, A\&A, 636, A113

\bibitem[{Gilli {et~al.}(2007)Gilli, Comastri, \& Hasinger}]{Gilli_2007}
Gilli, R., Comastri, A., \& Hasinger, G. 2007, A\&A, 463, 79

\bibitem[{{Goldreich} \& {Julian}(1969)}]{Goldreich_1969}
{Goldreich}, P. \& {Julian}, W.~H. 1969, \apj, 157, 869

\bibitem[{{H.E.S.S. Collaboration} {et~al.}(2019){H.E.S.S. Collaboration}, Abdalla, Aharonian, Ait~Benkhali, Angüner, {Arakawa, M.}, {Arcaro, C.}, {Armand, C.}, {Arrieta, M.}, {Backes, M.}, {Barnard, M.}, {Becherini, Y.}, {Becker Tjus, J.}, {Berge, D.}, {Bernlöhr, K.}, {Blackwell, R.}, {Böttcher, M.}, {Boisson, C.}, {Bolmont, J.}, {Bonnefoy, S.}, {Bordas, P.}, {Bregeon, J.}, {Brun, F.}, {Brun, P.}, {Bryan, M.}, {Büchele, M.}, {Bulik, T.}, {Bylund, T.}, {Capasso, M.}, {Caroff, S.}, {Carosi, A.}, {Casanova, S.}, {Cerruti, M.}, {Chakraborty, N.}, {Chand, T.}, {Chandra, S.}, {Chaves, R. C. G.}, {Chen, A.}, {Colafrancesco, S.}, {Condon, B.}, {Davids, I. D.}, {Deil, C.}, {Devin, J.}, {deWilt, P.}, {Dirson, L.}, {Djannati-Ataï, A.}, {Dmytriiev, A.}, {Donath, A.}, {Doroshenko, V.}, {Drury, L. O’C.}, {Dyks, J.}, {Egberts, K.}, {Emery, G.}, {Ernenwein, J.-P.}, {Eschbach, S.}, {Fegan, S.}, {Fiasson, A.}, {Fontaine, G.}, {Funk, S.}, {Füßling, M.}, {Gabici, S.}, {Gallant, Y. A.}, {Gaté, F.}, {Giavitto, G.},
  {Glawion, D.}, {Glicenstein, J. F.}, {Gottschall, D.}, {Grondin, M.-H.}, {Hahn, J.}, {Haupt, M.}, {Heinzelmann, G.}, {Henri, G.}, {Hermann, G.}, {Hinton, J. A.}, {Hofmann, W.}, {Hoischen, C.}, {Holch, T. L.}, {Holler, M.}, {Horns, D.}, {Huber, D.}, {Iwasaki, H.}, {Jacholkowska, A.}, {Jamrozy, M.}, {Jankowsky, D.}, {Jankowsky, F.}, {Jouvin, L.}, {Jung-Richardt, I.}, {Kastendieck, M. A.}, {Katarzyński, K.}, {Katsuragawa, M.}, {Katz, U.}, {Kerszberg, D.}, {Khangulyan, D.}, {Khélifi, B.}, {King, J.}, {Klepser, S.}, {Kluźniak, W.}, {Komin, Nu.}, {Kosack, K.}, {Kraus, M.}, {Lamanna, G.}, {Lau, J.}, {Lefaucheur, J.}, {Lemière, A.}, {Lemoine-Goumard, M.}, {Lenain, J.-P.}, {Leser, E.}, {Lohse, T.}, {López-Coto, R.}, {Lypova, I.}, {Malyshev, D.}, {Marandon, V.}, {Marcowith, A.}, {Mariaud, C.}, {Martí-Devesa, G.}, {Marx, R.}, {Maurin, G.}, {Meintjes, P. J.}, {Mitchell, A. M. W.}, {Moderski, R.}, {Mohamed, M.}, {Mohrmann, L.}, {Moore, C.}, {Moulin, E.}, {Murach, T.}, {Nakashima, S.}, {de Naurois, M.}, {Ndiyavala,
  H.}, {Niederwanger, F.}, {Niemiec, J.}, {Oakes, L.}, {O’Brien, P.}, {Odaka, H.}, {Ohm, S.}, {Ostrowski, M.}, {Oya, I.}, {Panter, M.}, {Parsons, R. D.}, {Perennes, C.}, {Petrucci, P.-O.}, {Peyaud, B.}, {Piel, Q.}, {Pita, S.}, {Poireau, V.}, {Priyana Noel, A.}, {Prokhorov, D. A.}, {Prokoph, H.}, {Pühlhofer, G.}, {Punch, M.}, {Quirrenbach, A.}, {Raab, S.}, {Rauth, R.}, {Reimer, A.}, {Reimer, O.}, {Renaud, M.}, {Rieger, F.}, {Rinchiuso, L.}, {Romoli, C.}, {Rowell, G.}, {Rudak, B.}, {Ruiz-Velasco, E.}, {Sahakian, V.}, {Saito, S.}, {Sanchez, D. A.}, {Santangelo, A.}, {Sasaki, M.}, {Schlickeiser, R.}, {Schüssler, F.}, {Schulz, A.}, {Schutte, H.}, {Schwanke, U.}, {Schwemmer, S.}, {Seglar-Arroyo, M.}, {Senniappan, M.}, {Seyffert, A. S.}, {Shafi, N.}, {Shilon, I.}, {Shiningayamwe, K.}, {Simoni, R.}, {Sinha, A.}, {Sol, H.}, {Specovius, A.}, {Spir-Jacob, M.}, {Stawarz, Ł.}, {Steenkamp, R.}, {Stegmann, C.}, {Steppa, C.}, {Takahashi, T.}, {Tavernet, J.-P.}, {Tavernier, T.}, {Taylor, A. M.}, {Terrier, R.}, {Tibaldo,
  L.}, {Tiziani, D.}, {Tluczykont, M.}, {Trichard, C.}, {Tsirou, M.}, {Tsuji, N.}, {Tuffs, R.}, {Uchiyama, Y.}, {van der Walt, D. J.}, {van Eldik, C.}, {van Rensburg, C.}, {van Soelen, B.}, {Vasileiadis, G.}, {Veh, J.}, {Venter, C.}, {Vincent, P.}, {Vink, J.}, {Voisin, F.}, {Völk, H. J.}, {Vuillaume, T.}, {Wadiasingh, Z.}, {Wagner, S. J.}, {Wagner, R. M.}, {White, R.}, {Wierzcholska, A.}, {Yang, R.}, {Yoneda, H.}, {Zaborov, D.}, {Zacharias, M.}, {Zanin, R.}, {Zdziarski, A. A.}, {Zech, A.}, {Zefi, F.}, {Ziegler, A.}, {Zorn, J.}, \& {Żywucka, N.}}]{HESSJ1825_2019}
{H.E.S.S. Collaboration}, Abdalla, H., Aharonian, F., {et~al.} 2019, A\&A, 621, A116

\bibitem[{{H.E.S.S. Collaboration} {et~al.}(2012){H.E.S.S. Collaboration}, Abramowski, Acero, Aharonian, Akhperjanian, {Anton, G.}, {Balenderan, S.}, {Balzer, A.}, {Barnacka, A.}, {Becherini, Y.}, {Becker, J.}, {Bernlöhr, K.}, {Birsin, E.}, {Biteau, J.}, {Bochow, A.}, {Boisson, C.}, {Bolmont, J.}, {Bordas, P.}, {Brucker, J.}, {Brun, F.}, {Brun, P.}, {Bulik, T.}, {Büsching, I.}, {Carrigan, S.}, {Casanova, S.}, {Cerruti, M.}, {Chadwick, P. M.}, {Charbonnier, A.}, {Chaves, R. C. G.}, {Cheesebrough, A.}, {Cologna, G.}, {Conrad, J.}, {Couturier, C.}, {Dalton, M.}, {Daniel, M. K.}, {Davids, I. D.}, {Degrange, B.}, {Deil, C.}, {Dickinson, H. J.}, {Djannati-Ataï, A.}, {Domainko, W.}, {Drury, L. O’C.}, {Dubus, G.}, {Dutson, K.}, {Dyks, J.}, {Dyrda, M.}, {Egberts, K.}, {Eger, P.}, {Espigat, P.}, {Fallon, L.}, {Farnier, C.}, {Fegan, S.}, {Feinstein, F.}, {Fernandes, M. V.}, {Fiasson, A.}, {Fontaine, G.}, {Förster, A.}, {Füßling, M.}, {Gajdus, M.}, {Gallant, Y. A.}, {Garrigoux, T.}, {Gast, H.}, {Gérard, L.},
  {Giebels, B.}, {Glicenstein, J. F.}, {Glück, B.}, {Göring, D.}, {Grondin, M.-H.}, {Häffner, S.}, {Hague, J. D.}, {Hahn, J.}, {Hampf, D.}, {Harris, J.}, {Hauser, M.}, {Heinz, S.}, {Heinzelmann, G.}, {Henri, G.}, {Hermann, G.}, {Hillert, A.}, {Hinton, J. A.}, {Hofmann, W.}, {Hofverberg, P.}, {Holler, M.}, {Horns, D.}, {Jacholkowska, A.}, {Jahn, C.}, {Jamrozy, M.}, {Jung, I.}, {Kastendieck, M. A.}, {Katarzyński, K.}, {Katz, U.}, {Kaufmann, S.}, {Khélifi, B.}, {Klochkov, D.}, {Kluźniak, W.}, {Kneiske, T.}, {Komin, Nu.}, {Kosack, K.}, {Kossakowski, R.}, {Krayzel, F.}, {Laffon, H.}, {Lamanna, G.}, {Lenain, J.-P.}, {Lennarz, D.}, {Lohse, T.}, {Lopatin, A.}, {Lu, C.-C.}, {Marandon, V.}, {Marcowith, A.}, {Masbou, J.}, {Maurin, G.}, {Maxted, N.}, {Mayer, M.}, {McComb, T. J. L.}, {Medina, M. C.}, {Méhault, J.}, {Menzler, U.}, {Moderski, R.}, {Mohamed, M.}, {Moulin, E.}, {Naumann, C. L.}, {Naumann-Godo, M.}, {de Naurois, M.}, {Nedbal, D.}, {Nekrassov, D.}, {Nguyen, N.}, {Nicholas, B.}, {Niemiec, J.}, {Nolan, S.
  J.}, {Ohm, S.}, {de Oña Wilhelmi, E.}, {Opitz, B.}, {Ostrowski, M.}, {Oya, I.}, {Panter, M.}, {Paz Arribas, M.}, {Pekeur, N. W.}, {Pelletier, G.}, {Perez, J.}, {Petrucci, P.-O.}, {Peyaud, B.}, {Pita, S.}, {Pühlhofer, G.}, {Punch, M.}, {Quirrenbach, A.}, {Raue, M.}, {Reimer, A.}, {Reimer, O.}, {Renaud, M.}, {de los Reyes, R.}, {Rieger, F.}, {Ripken, J.}, {Rob, L.}, {Rosier-Lees, S.}, {Rowell, G.}, {Rudak, B.}, {Rulten, C. B.}, {Sahakian, V.}, {Sanchez, D. A.}, {Santangelo, A.}, {Schlickeiser, R.}, {Schulz, A.}, {Schwanke, U.}, {Schwarzburg, S.}, {Schwemmer, S.}, {Sheidaei, F.}, {Skilton, J. L.}, {Sol, H.}, {Spengler, G.}, {Stawarz, Ł.}, {Steenkamp, R.}, {Stegmann, C.}, {Stinzing, F.}, {Stycz, K.}, {Sushch, I.}, {Szostek, A.}, {Tavernet, J.-P.}, {Terrier, R.}, {Tluczykont, M.}, {Valerius, K.}, {van Eldik, C.}, {Vasileiadis, G.}, {Venter, C.}, {Viana, A.}, {Vincent, P.}, {Völk, H. J.}, {Volpe, F.}, {Vorobiov, S.}, {Vorster, M.}, {Wagner, S. J.}, {Ward, M.}, {White, R.}, {Wierzcholska, A.}, {Zacharias, M.},
  {Zajczyk, A.}, {Zdziarski, A. A.}, {Zech, A.}, \& {Zechlin, H.-S.}}]{HESS_J1303}
{H.E.S.S. Collaboration}, Abramowski, A., Acero, F., {et~al.} 2012, A\&A, 548, A46

\bibitem[{{H.E.S.S. Collaboration} {et~al.}(2011){H.E.S.S. Collaboration}, Abramowski, Acero, Aharonian, Akhperjanian, {Anton, G.}, {Balzer, A.}, {Barnacka, A.}, {Barres de Almeida, U.}, {Becherini, Y.}, {Becker, J.}, {Behera, B.}, {Bernlöhr, K.}, {Bochow, A.}, {Boisson, C.}, {Bolmont, J.}, {Bordas, P.}, {Brucker, J.}, {Brun, F.}, {Brun, P.}, {Bulik, T.}, {Büsching, I.}, {Carrigan, S.}, {Casanova, S.}, {Cerruti, M.}, {Chadwick, P. M.}, {Charbonnier, A.}, {Chaves, R. C. G.}, {Cheesebrough, A.}, {Chounet, L.-M.}, {Clapson, A. C.}, {Coignet, G.}, {Cologna, G.}, {Conrad, J.}, {Dalton, M.}, {Daniel, M. K.}, {Davids, I. D.}, {Degrange, B.}, {Deil, C.}, {Dickinson, H. J.}, {Djannati-Ataï, A.}, {Domainko, W.}, {Drury, L. O’C.}, {Dubois, F.}, {Dubus, G.}, {Dutson, K.}, {Dyks, J.}, {Dyrda, M.}, {Egberts, K.}, {Eger, P.}, {Espigat, P.}, {Fallon, L.}, {Farnier, C.}, {Fegan, S.}, {Feinstein, F.}, {Fernandes, M. V.}, {Fiasson, A.}, {Fontaine, G.}, {Förster, A.}, {Füßling, M.}, {Gallant, Y. A.}, {Gast, H.},
  {Gérard, L.}, {Gerbig, D.}, {Giebels, B.}, {Glicenstein, J. F.}, {Glück, B.}, {Goret, P.}, {Göring, D.}, {Häffner, S.}, {Hague, J. D.}, {Hampf, D.}, {Hauser, M.}, {Heinz, S.}, {Heinzelmann, G.}, {Henri, G.}, {Hermann, G.}, {Hinton, J. A.}, {Hoffmann, A.}, {Hofmann, W.}, {Hofverberg, P.}, {Holler, M.}, {Horns, D.}, {Jacholkowska, A.}, {de Jager, O. C.}, {Jahn, C.}, {Jamrozy, M.}, {Jung, I.}, {Kastendieck, M. A.}, {Katarzyński, K.}, {Katz, U.}, {Kaufmann, S.}, {Keogh, D.}, {Khangulyan, D.}, {Khélifi, B.}, {Klochkov, D.}, {Kluźniak, W.}, {Kneiske, T.}, {Komin, Nu.}, {Kosack, K.}, {Kossakowski, R.}, {Laffon, H.}, {Lamanna, G.}, {Lennarz, D.}, {Lohse, T.}, {Lopatin, A.}, {Lu, C.-C.}, {Marandon, V.}, {Marcowith, A.}, {Masbou, J.}, {Maurin, D.}, {Maxted, N.}, {Mayer, M.}, {McComb, T. J. L.}, {Medina, M. C.}, {Méhault, J.}, {Moderski, R.}, {Moulin, E.}, {Naumann, C. L.}, {Naumann-Godo, M.}, {de Naurois, M.}, {Nedbal, D.}, {Nekrassov, D.}, {Nguyen, N.}, {Nicholas, B.}, {Niemiec, J.}, {Nolan, S. J.}, {Ohm,
  S.}, {de Oña Wilhelmi, E.}, {Opitz, B.}, {Ostrowski, M.}, {Oya, I.}, {Panter, M.}, {Paz Arribas, M.}, {Pedaletti, G.}, {Pelletier, G.}, {Petrucci, P.-O.}, {Pita, S.}, {Pühlhofer, G.}, {Punch, M.}, {Quirrenbach, A.}, {Raue, M.}, {Rayner, S. M.}, {Reimer, A.}, {Reimer, O.}, {Renaud, M.}, {de los Reyes, R.}, {Rieger, F.}, {Ripken, J.}, {Rob, L.}, {Rosier-Lees, S.}, {Rowell, G.}, {Rudak, B.}, {Rulten, C. B.}, {Ruppel, J.}, {Sahakian, V.}, {Sanchez, D.}, {Santangelo, A.}, {Schlickeiser, R.}, {Schöck, F. M.}, {Schulz, A.}, {Schwanke, U.}, {Schwarzburg, S.}, {Schwemmer, S.}, {Sikora, M.}, {Skilton, J. L.}, {Sol, H.}, {Spengler, G.}, {Stawarz, Ł.}, {Steenkamp, R.}, {Stegmann, C.}, {Stinzing, F.}, {Stycz, K.}, {Sushch, I.}, {Szostek, A.}, {Tavernet, J.-P.}, {Terrier, R.}, {Tluczykont, M.}, {Valerius, K.}, {van Eldik, C.}, {Vasileiadis, G.}, {Venter, C.}, {Vialle, J. P.}, {Viana, A.}, {Vincent, P.}, {Völk, H. J.}, {Volpe, F.}, {Vorobiov, S.}, {Vorster, M.}, {Wagner, S. J.}, {Ward, M.}, {White, R.},
  {Wierzcholska, A.}, {Zacharias, M.}, {Zajczyk, A.}, {Zdziarski, A. A.}, {Zech, A.}, \& {Zechlin, H.-S.}}]{HESS_J1356}
{H.E.S.S. Collaboration}, Abramowski, A., Acero, F., {et~al.} 2011, A\&A, 533, A103

\bibitem[{{H.E.S.S. Collaboration} {et~al.}(2023){H.E.S.S. Collaboration}, Aharonian, Ait~Benkhali, Aschersleben, Ashkar, Backes, Barbosa~Martins, {Batzofin}, {Becherini}, {Berge}, {Bernl{\"o}hr}, {Bi}, {B{\"o}ttcher}, {Boisson}, {Bolmont}, {de Bony de Lavergne}, {Borowska}, {Bradascio}, {Breuhaus}, {Brose}, {Brun}, {Bruno}, {Bulik}, {Burger-Scheidlin}, {Bylund}, {Cangemi}, {Caroff}, {Casanova}, {Celic}, {Cerruti}, {Chand}, {Chandra}, {Chen}, {Chibueze}, {Cotter}, {Damascene Mbarubucyeye}, {Djannati-Ata{\"\i}}, {Dmytriiev}, {Egberts}, {Ernenwein}, {Feijen}, {Fiasson}, {Fichet de Clairfontaine}, {Fontaine}, {F{\"u}{\ss}ling}, {Funk}, {Gabici}, {Gallant}, {Ghafourizadeh}, {Giavitto}, {Giunti}, {Glawion}, {Glicenstein}, {Goswami}, {Grolleron}, {Grondin}, {Haerer}, {Haupt}, {Hinton}, {Hofmann}, {Holch}, {Holler}, {Horns}, {Huang}, {Jamrozy}, {Jankowsky}, {Joshi}, {Jung-Richardt}, {Kasai}, {Katarzy{\'n}ski}, {Kh{\'e}lifi}, {Klepser}, {Klu{\v{z}}niak}, {Komin}, {Kosack}, {Kostunin}, {Lang}, {Le Stum},
  {Lemi{\`e}re}, {Lemoine-Goumard}, {Lenain}, {Leuschner}, {Lohse}, {Luashvili}, {Lypova}, {Mackey}, {Malyshev}, {Malyshev}, {Marandon}, {Marchegiani}, {Marcowith}, {Marinos}, {Mart{\'\i}-Devesa}, {Marx}, {Maurin}, {Meyer}, {Mitchell}, {Moderski}, {Mohrmann}, {Montanari}, {Moulin}, {Muller}, {Murach}, {Nakashima}, {de Naurois}, {Niemiec}, {Noel}, {O'Brien}, {Ohm}, {Olivera-Nieto}, {de Ona Wilhelmi}, {Ostrowski}, {Panny}, {Panter}, {Parsons}, {Peron}, {Pita}, {Prokhorov}, {Prokoph}, {P{\"u}hlhofer}, {Punch}, {Quirrenbach}, {Reichherzer}, {Reimer}, {Reimer}, {Renaud}, {Rieger}, {Rowell}, {Rudak}, {Ruiz-Velasco}, {Sahakian}, {Sailer}, {Salzmann}, {Sanchez}, {Santangelo}, {Sasaki}, {Sch{\"u}ssler}, {Schwanke}, {Shapopi}, {Sinha}, {Sol}, {Specovius}, {Spencer}, {Spir-Jacob}, {Stawarz}, {Steenkamp}, {Steinmassl}, {Steppa}, {Sushch}, {Suzuki}, {Takahashi}, {Tanaka}, {Tavernier}, {Terrier}, {Thorpe-Morgan}, {Tluczykont}, {Tsirou}, {Tsuji}, {van Eldik}, {Vecchi}, {Veh}, {Venter}, {Vink}, {Wagner}, {Werner}, {White},
  {Wierzcholska}, {Wun Wong}, {Yassin}, {Zacharias}, {Zargaryan}, {Zdziarski}, {Zech}, {Zhu}, {Zouari}, {{\.Z}ywucka}, {Zanin}, {Kerr}, {Johnston}, {Shannon}, \& {Smith}}]{HESS_2023}
{H.E.S.S. Collaboration}, Aharonian, F., Ait~Benkhali, F., {et~al.} 2023, Nature Astronomy, 7, 1341

\bibitem[{Iwashita {et~al.}(2023)Iwashita, Kataoka, \& Sofue}]{Iwashita_2023}
Iwashita, R., Kataoka, J., \& Sofue, Y. 2023, \apj, 958, 83

\bibitem[{Kargaltsev {et~al.}(2008{\natexlab{a}})Kargaltsev, Misanovic, Pavlov, Wong, \& Garmire}]{Kargaltsev_2008}
Kargaltsev, O., Misanovic, Z., Pavlov, G.~G., Wong, J.~A., \& Garmire, G.~P. 2008{\natexlab{a}}, \apj, 684, 542

\bibitem[{Kargaltsev \& Pavlov(2008)}]{Kargaltsev_Pavlov_2008}
Kargaltsev, O. \& Pavlov, G.~G. 2008, in American Institute of Physics Conference Series, Vol. 983, 40 Years of Pulsars: Millisecond Pulsars, Magnetars and More, ed. C.~{Bassa}, Z.~{Wang}, A.~{Cumming}, \& V.~M. {Kaspi} (AIP), 171--185

\bibitem[{Kargaltsev \& Pavlov(2010)}]{Kargaltsev_Pavlov_2010}
Kargaltsev, O. \& Pavlov, G.~G. 2010, AIP Conference Proceedings, 1248, 25

\bibitem[{Kargaltsev {et~al.}(2017)Kargaltsev, Pavlov, Klingler, \& Rangelov}]{Kargaltsev_2017}
Kargaltsev, O., Pavlov, G.~G., Klingler, N., \& Rangelov, B. 2017, Journal of Plasma Physics, 83, 635830501

\bibitem[{Kargaltsev {et~al.}(2008{\natexlab{b}})Kargaltsev, Pavlov, \& Wong}]{Kargaltsev_2009}
Kargaltsev, O., Pavlov, G.~G., \& Wong, J.~A. 2008{\natexlab{b}}, ApJ, 690, 891

\bibitem[{Kataoka {et~al.}(2018)Kataoka, Sofue, Inoue, Akita, Nakashima, \& Totani}]{Kataoka_2018}
Kataoka, J., Sofue, Y., Inoue, Y., {et~al.} 2018, Galaxies, 6

\bibitem[{Khangulyan {et~al.}(2014)Khangulyan, Aharonian, \& Kelner}]{Khangulyan_2014}
Khangulyan, D., Aharonian, F.~A., \& Kelner, S.~R. 2014, \apj, 783, 100

\bibitem[{Khokhriakova {et~al.}(2024)Khokhriakova, Becker, Ponti, Sasaki, Li, \& Liu}]{Khokhriakova_2024}
Khokhriakova, A., Becker, W., Ponti, G., {et~al.} 2024, A\&A, 683, A180

\bibitem[{Khokhriakova {et~al.}(2025)Khokhriakova, Becker, Predehl, Sanders, Freyberg, \& Schwope}]{Khokhriakova_2025}
Khokhriakova, A., Becker, W., Predehl, P., {et~al.} 2025, Research Notes of the AAS, 9, 154

\bibitem[{Klingler {et~al.}(2018)Klingler, Kargaltsev, Pavlov, Ng, Beniamini, \& Volkov}]{Klingler_2018}
Klingler, N., Kargaltsev, O., Pavlov, G.~G., {et~al.} 2018, \apj, 861, 5

\bibitem[{Klingler {et~al.}(2016)Klingler, Rangelov, Kargaltsev, Pavlov, Romani, Posselt, Slane, Temim, Ng, Bucciantini, Bykov, Swartz, \& Buehler}]{Klingler_2016}
Klingler, N., Rangelov, B., Kargaltsev, O., {et~al.} 2016, \apj, 833, 253

\bibitem[{Kuntz \& Snowden(2008)}]{Kuntz_Snowden_2008}
Kuntz, K.~D. \& Snowden, S.~L. 2008, A\&A, 478, 575

\bibitem[{Leccardi \& Molendi(2008)}]{Leccardi_Molendi_2008}
Leccardi, A. \& Molendi, S. 2008, A\&A, 486, 359

\bibitem[{{LHAASO Collaboration} {et~al.}(2021){LHAASO Collaboration}, Cao, Aharonian, An, null, Bai, Bai, Bao, Bastieri, Bi, Bi, Cai, Cai, Cao, Chang, Chang, Chen, Chen, Chen, Chen, Chen, Chen, Chen, Chen, Chen, Chen, Chen, Chen, Chen, Chen, Cheng, Cheng, Cui, Cui, Cui, Piazzoli, Dai, Dai, Dai, null, della Volpe, Dong, Duan, Fan, Fan, Fan, Fang, Fang, Feng, Feng, Feng, Feng, Gao, Gao, Gao, Gao, Gao, Ge, Geng, Gong, Gou, Gu, Guo, Guo, Guo, Guo, Guo, Han, He, He, He, He, He, He, Heller, Hor, Hou, Hou, Hu, Hu, Hu, Hu, Huang, Huang, Huang, Huang, Huang, Huang, Ji, Ji, Jia, Jiang, Jiang, Jin, Ke, Kuleshov, Levochkin, Li, Li, Li, Li, Li, Li, Li, Li, Li, Li, Li, Li, Li, Li, Li, Li, Li, Li, Liang, Liang, Lin, Liu, Liu, Liu, Liu, Liu, Liu, Liu, Liu, Liu, Liu, Liu, Liu, Liu, Liu, Liu, Liu, Long, Lu, Lv, Ma, Ma, Ma, Mao, Masood, Min, Mitthumsiri, Montaruli, Nan, Pang, Pattarakijwanich, Pei, Qi, Qi, Qiao, Qin, Ruffolo, Rulev, Saiz, Shao, Shchegolev, Sheng, Shi, Song, Stenkin, Stepanov, Su, Sun, Sun, Sun, Tam, Tang,
  Tian, Wang, Wang, Wang, Wang, Wang, Wang, Wang, Wang, Wang, Wang, Wang, Wang, Wang, Wang, Wang, Wang, Wang, Wang, Wang, Wang, Wang, Wang, Wei, Wei, Wei, Wen, Wu, Wu, Wu, Wu, Wu, Xi, Xia, Xia, Xiang, Xiao, Xiao, Xiao, Xin, Xin, Xing, Xu, Xu, Xue, Yan, Yan, Yang, Yang, Yang, Yang, Yang, Yang, Yang, Yao, Yao, Ye, Yin, Yin, You, You, Yu, Yuan, Zeng, Zeng, Zeng, Zeng, Zha, Zhai, Zhang, Zhang, Zhang, Zhang, Zhang, Zhang, Zhang, Zhang, Zhang, Zhang, Zhang, Zhang, Zhang, Zhang, Zhang, Zhang, Zhang, Zhang, Zhang, Zhao, Zhao, Zhao, Zhao, Zhao, Zheng, Zheng, Zhou, Zhou, Zhou, Zhou, Zhou, Zhou, Zhu, Zhu, Zhu, Zhu, \& Zuo}]{LHAASO_crab_2021}
{LHAASO Collaboration}, Cao, Z., Aharonian, F., {et~al.} 2021, Science, 373, 425

\bibitem[{{LHAASO Collaboration} {et~al.}(2025){LHAASO Collaboration}, Cao, {et~al.}}]{LHAASO_2025}
{LHAASO Collaboration}, Cao, Z., {et~al.} 2025, The Innovation, 100802

\bibitem[{Li {et~al.}(2022)Li, Zhang, Liu, Liu, \& Wang}]{Li_2021}
Li, B., Zhang, Y., Liu, T., Liu, R.-Y., \& Wang, X.-Y. 2022, \mnras, 513, 2884

\bibitem[{Liu {et~al.}(2019{\natexlab{a}})Liu, Ge, Sun, \& Wang}]{Liu_2019}
Liu, R.-Y., Ge, C., Sun, X.-N., \& Wang, X.-Y. 2019{\natexlab{a}}, \apj, 875, 149

\bibitem[{Liu {et~al.}(2019{\natexlab{b}})Liu, Yan, \& Zhang}]{Liu_Yan_2019}
Liu, R.-Y., Yan, H., \& Zhang, H. 2019{\natexlab{b}}, \prl, 123, 221103

\bibitem[{Locatelli {et~al.}(2022)Locatelli, Ponti, \& Bianchi}]{Locatelli_2022}
Locatelli, N., Ponti, G., \& Bianchi, S. 2022, \aap, 659, A118

\bibitem[{L{\'o}pez-Coto {et~al.}(2022)L{\'o}pez-Coto, de~O{\~n}a~Wilhelmi, Aharonian, Amato, \& Hinton}]{LopezCoto_2022}
L{\'o}pez-Coto, R., de~O{\~n}a~Wilhelmi, E., Aharonian, F., Amato, E., \& Hinton, J. 2022, Nature Astronomy, 6, 199

\bibitem[{Manchester {et~al.}(2005)Manchester, Hobbs, Teoh, \& Hobbs}]{Manchester_2005}
Manchester, R.~N., Hobbs, G.~B., Teoh, A., \& Hobbs, M. 2005, The Astronomical Journal, 129, 1993

\bibitem[{Manconi {et~al.}(2024)Manconi, Woo, Shang, Krivonos, Tang, Di~Mauro, Donato, Mori, \& Hailey}]{Manconi_2024}
Manconi, S., Woo, J., Shang, R.-Y., {et~al.} 2024, A\&A, 689, A326

\bibitem[{McLaughlin {et~al.}(2000)McLaughlin, Cordes, \& Arzoumanian}]{McLaughlin_2000}
McLaughlin, M., Cordes, J.~M., \& Arzoumanian, Z. 2000, in Astronomical Society of the Pacific Conference Series, Vol. 202, IAU Colloq. 177: Pulsar Astronomy - 2000 and Beyond, ed. M.~{Kramer}, N.~{Wex}, \& R.~{Wielebinski}, 41

\bibitem[{McLaughlin {et~al.}(2002)McLaughlin, Arzoumanian, Cordes, Backer, Lommen, Lorimer, \& Zepka}]{McLaughlin_2002}
McLaughlin, M.~A., Arzoumanian, Z., Cordes, J.~M., {et~al.} 2002, \apj, 564, 333

\bibitem[{{Niu} {et~al.}(2025){Niu}, {Yuan}, {Zhang}, {Lei}, {Ji}, \& {Fan}}]{Niu_2025}
{Niu}, S., {Yuan}, Q., {Zhang}, S.-N., {et~al.} 2025, arXiv e-prints, arXiv:2501.17046

\bibitem[{Olmi(2023)}]{Olmi_2023}
Olmi, B. 2023, Universe, 9, 402

\bibitem[{Olmi \& Bucciantini(2019)}]{Olmi_Bucciantini_2019}
Olmi, B. \& Bucciantini, N. 2019, \mnras, 484, 5755

\bibitem[{Pavan {et~al.}(2016)Pavan, Pühlhofer, Bordas, Audard, Balbo, Bozzo, Eckert, Ferrigno, Filipović, Verdugo, \& Walter}]{Pavan_2016}
Pavan, L., Pühlhofer, G., Bordas, P., {et~al.} 2016, A\&A, 591, A91

\bibitem[{Pellizzoni {et~al.}(2004)Pellizzoni, Mattana, Mereghetti, De~Luca, Caraveo, \& Conti}]{Pellizzoni_2004}
Pellizzoni, A., Mattana, F., Mereghetti, S., {et~al.} 2004, Memorie della Societa Astronomica Italiana Supplementi, 5, 195

\bibitem[{Ponti {et~al.}(2023)Ponti, Zheng, Locatelli, Bianchi, Zhang, Anastasopoulou, Comparat, Dennerl, Freyberg, Haberl, Merloni, Reiprich, Salvato, Sanders, Sasaki, Strong, \& Yeung}]{Ponti_2023}
Ponti, G., Zheng, X., Locatelli, N., {et~al.} 2023, A\&A, 674, A195

\bibitem[{Popescu {et~al.}(2017)Popescu, Yang, Tuffs, Natale, Rushton, \& Aharonian}]{Popescu_2017}
Popescu, C.~C., Yang, R., Tuffs, R.~J., {et~al.} 2017, \mnras, 470, 2539

\bibitem[{Rees \& Gunn(1974)}]{Rees_1974}
Rees, M. \& Gunn, J.~E. 1974, \mnras, 167, 1

\bibitem[{Rigoselli {et~al.}(2022)Rigoselli, Mereghetti, Anzuinelli, Keith, Taverna, Turolla, \& Zane}]{Rigoselli_2022}
Rigoselli, M., Mereghetti, S., Anzuinelli, S., {et~al.} 2022, \mnras, 513, 3113

\bibitem[{Schwarz(1978)}]{Schwarz_1978}
Schwarz, G. 1978, The Annals of Statistics, 6, 461

\bibitem[{Sudoh {et~al.}(2019)Sudoh, Linden, \& Beacom}]{Sudoh_2019}
Sudoh, T., Linden, T., \& Beacom, J.~F. 2019, Phys. Rev. D, 100, 043016

\bibitem[{{The VERITAS Collaboration} {et~al.}(2011){The VERITAS Collaboration}, Aliu, Arlen, Aune, Beilicke, Benbow, Bouvier, Bradbury, Buckley, Bugaev, Byrum, Cannon, Cesarini, Christiansen, Ciupik, Collins-Hughes, Connolly, Cui, Dickherber, Duke, Errando, Falcone, Finley, Finnegan, Fortson, Furniss, Galante, Gall, Gibbs, Gillanders, Godambe, Griffin, Grube, Guenette, Gyuk, Hanna, Holder, Huan, Hughes, Hui, Humensky, Imran, Kaaret, Karlsson, Kertzman, Kieda, Krawczynski, Krennrich, Lang, Lyutikov, Madhavan, Maier, Majumdar, McArthur, McCann, McCutcheon, Moriarty, Mukherjee, Nuñez, Ong, Orr, Otte, Park, Perkins, Pizlo, Pohl, Prokoph, Quinn, Ragan, Reyes, Reynolds, Roache, Rose, Ruppel, Saxon, Schroedter, Sembroski, Şentürk, Smith, Staszak, Tešić, Theiling, Thibadeau, Tsurusaki, Tyler, Varlotta, Vassiliev, Vincent, Vivier, Wakely, Ward, Weekes, Weinstein, Weisgarber, Williams, \& Zitzer}]{VERITAS_2011}
{The VERITAS Collaboration}, Aliu, E., Arlen, T., {et~al.} 2011, Science, 334, 69

\bibitem[{van~der Swaluw {et~al.}(2004)van~der Swaluw, Downes, \& Keegan}]{van_der_Swaluw_2004}
van~der Swaluw, E., Downes, T.~P., \& Keegan, R. 2004, A\&A, 420, 937

\bibitem[{Van~Etten \& Romani(2011)}]{VanEtten_2011}
Van~Etten, A. \& Romani, R.~W. 2011, \apj, 742, 62

\bibitem[{Wilms {et~al.}(2000)Wilms, Allen, \& McCray}]{Wilms_2000}
Wilms, J., Allen, A., \& McCray, R. 2000, \apj, 542, 914

\bibitem[{{Xie} {et~al.}(2025){Xie}, {Liu}, {Shao}, {Cui}, \& {Yang}}]{Xie_2025}
{Xie}, C., {Liu}, Y., {Shao}, C., {Cui}, Y., \& {Yang}, L. 2025, \aap, 698, A98

\bibitem[{Yao {et~al.}(2017)Yao, Manchester, \& Wang}]{Yao_2017}
Yao, J.~M., Manchester, R.~N., \& Wang, N. 2017, \apj, 835, 29

\bibitem[{Zabalza(2015)}]{naima}
Zabalza, V. 2015, PoS - ICRC2015, 922

\bibitem[{Zhang {et~al.}(2024)Zhang, Ponti, Carretti, Liu, Morris, {Haverkorn}, {Locatelli}, {Zheng}, {Aharonian}, {Zhang}, {Zhang}, {Stel}, {Strong}, {Yeung}, \& {Merloni}}]{Zhang_2024}
Zhang, H.-S., Ponti, G., Carretti, E., {et~al.} 2024, Nature Astronomy, 8, 1416

\bibitem[{Zirakashvili \& Aharonian(2007)}]{Zirakashvili_2007}
Zirakashvili, V.~N. \& Aharonian, F. 2007, \aap, 465, 695

\end{thebibliography}

\end{document}